**Memoryless quantum repeaters based on cavity-QED and coherent states**

*Pei-Zhe Li and Peter van Loock\**

P.-Z. Li,
Department of Informatics, School of Multidisciplinary Sciences, SOKENDAI (The Graduate University for Advanced Studies), 2-1-2 Hitotsubashi, Chiyoda-ku, Tokyo 101-8430, Japan
National Institute of Informatics, 2-1-2 Hitotsubashi, Chiyoda-ku, Tokyo 101-8430, Japan
Quantum Information Science and Technology Unit, Okinawa Institute of Science and Technology (OIST) Graduate University, Okinawa 904-0495, Japan
Prof. P. van Loock
Institut für Physik, Johannes Gutenberg-Universität, Staudingerweg 7, Mainz 55128, Germany
E-mail: loock@uni-mainz.de

A quantum repeater scheme based on cavity-QED and quantum error correction of channel loss via rotation-symmetric bosonic codes (**RSBC**) is proposed to distribute atomic entangled states over long distances without memories and at high clock rates. In this scheme, controlled rotation gates, i.e., phase shifts of the propagating light modes conditioned upon the state of an atom placed in a cavity, provide a mechanism both for the entangled-state preparations and for the error syndrome identifications. The distributed entangled pairs can then be used for quantum key distribution (**QKD**). In order to assess the performance of this repeater protocol, an explicit instance of RSBC—multi-component cat codes are studied quantitatively. A numerical simulation shows that the total fidelity and the success probability for quantum communication over a long distance (such as 1000km) both can almost approach unity provided a small enough elementary distance between stations (smaller than 0.1km or 0.01km) and rather low local losses (up to 0.1%) are considered. Secret key rates can become correspondingly high, both per channel use, beating the repeaterless bound, and per second thanks to the relatively high clock rates of the memoryless scheme. It is predicted from these results that a larger elementary distance might also keep both the fidelity and the success probability close to unity if higher-loss codes are employed. Based upon the cavity-QED setting, this scheme can be realized at room temperature and at optical frequencies.

## 1. Introduction

One of the key topics in quantum information processing recently is quantum communication over long distances, beyond what is achievable in a so-called point-to point link via direct optical state transmission. In order to obtain high fidelities and rates (or secret key rates in QKD), the exponential decrease of these quantities with distance, caused by the lossy bosonic channel (**LBC**) representing an optical fiber, must be suppressed. The most common and best known approach to remedy this fundamental scaling problem of fiber-based quantum communication is the so-called quantum repeater in which an otherwise long-distance point-to-point communication channel is divided into shorter segments.[1] The intermediate repeater stations are then equipped with stationary matter qubits that can couple to the incoming, flying photonic qubits, temporarily store and release them. Based on this, the typically heralded distribution of entangled photon pairs can be synchronized among all the repeater segments and these short-distance pairs can then be connected via quantum teleportation (entanglement swapping). In order for this scheme to be really scalable to larger distances, additional steps of entanglement distillation are needed to suppress the accumulation of memory and gate errors. This standard quantum repeater based on quantum memories and probabilistic entanglement distillation, though scalable in principle, has a couple of complications. While the requirement of an efficient, long-lasting quantum memory system is of more technical, experimental nature, a more conceptual and fundamental issue is the need for two-way classical communication in such repeaters, even over distances beyond those of the elementary segment lengths, significantly limiting the possible final rates. The use of quantum error correction codes for the matter qubits allows to circumvent the need for probabilistic operations on higher repeater levels and hence for any two-way classical communication between non-nearest repeater stations.[2] While this approach helps to enhance the final rates in principle, it still requires quantum memories and two-way classical communication for the initial encoded, entangled state distributions and distillations. Instead, a more recent concept would rather avoid the need for quantum memories and two-way classical communication entirely by minimizing the effect of the LBC through certain optical encodings, i.e. photonic or, more generally, bosonic quantum error correction codes that protect a bosonic mode from loss.[3] In this case, the clock rate of the repeater scheme is no longer determined by any classical-signal waiting times, but solely depends on the longest time duration required for the local processing at any repeater station. The motivation of this work is to propose and assess a similar, memoryless repeater scheme for long-distance quantum communication making use of 'Schrödinger cat states' for bosonic quantum error correction (QEC) in order to improve the fidelities and the rates.

The Cat codes,[4] is a prime example of a bosonic code where a logical qubit is encoded in an oscillator mode. Thus, it is sometimes also referred to as an instance of a continuous-variable (CV) code. It was proposed in particular to correct photon loss occurring on the bosonic mode exploiting certain photon number parity properties of suitable cat states. With the help of cavity-QED, cat states can be deterministically generated at room temperature in the laboratory.[5] Our work is especially based on the idea that the experimental scheme of Ref. [5] could not only be used to generate (atom-light entangled) cat states, but also to do the syndrome detections for cat-code error correction. Using this method, it is possible to construct a model of quantum repeaters (**QR**s) that do not require any memories or two-way classical communications. More generally, it is possible to extend the simplest instance of such a scheme to create multi-component cat states, which can be used for quantum error correction with higher-loss cat codes.[6,7] There also exists a more general class of quantum states and codes compared to that of multi-component cat states or codes. These more general codes are referred to as 'Rotation-Symmetric Bosonic Codes' (**RSBC**) and they exhibit, like the 'cat states', rotational symmetries in phase space.[8] We will argue that, in principle, all the above-mentioned operations for 'cat states' also work for the generalized class of states. First, we can physically prepare these states in the lab given that we are able to create a corresponding 'primitive state', which is useful in the context of quantum computing.[8] Secondly, these states are also potentially useful in long-distance quantum communication in the same way as the cat codes. Nonetheless, here our quantitative analysis will focus on cat codes and we shall not explicitly assess the performance of the generalized states, even though they could help to enhance the performance beyond that of cat codes. For the cat codes, however, we will systematically derive and quantitatively assess instances of the higher-order loss codes[6,7] spanned by more than those four distinct coherent states as for the original one-loss cat code.[4] While there are all-optical proposals for state generation and error correction with four coherent states,[9–12] it is unclear whether and how these are extendible to the regime of higher codes with more coherent states.

Compared with most existing schemes for fast, memoryless quantum communication or so-called 3$^{rd}$-generation quantum repeaters[3] where logical qubits are encoded in many physical, photonic qubits and twice as many optical modes, our scheme is particularly 'hardware-efficient' on the optical level by exploiting bosonic single-mode codes that are sent through the fiber channel. To some extent, this concept was already explored in Refs. [6,7], but an explicit optical realization has not been considered yet. Other existing approaches for fast, long-distance quantum communication rely upon a different, shift-invariant class of bosonic codes that also encode logical qubits into oscillator modes,[13] but partially also make use of concatenations with higher-level qubit codes.[14,15] An experimental complication in these schemes, however, is that it is even hard to create the corresponding low-level code states in the optical regime, although also for this proposals based on cavity-QED exist.[16,17] For logical qubits encoded into cat states, it may also be useful to protect the cat qubits through higher-level qubit quantum error correction codes.[11,18,19] While this can lead to a further improvement of the possible (secret key) rates, it would also render the state generations and encoding circuits more complicated and, by relying upon many optical modes for the total code states, it would go beyond the efficient bosonic single-mode encoding and transmission.

Our cat code repeater scheme, though functioning without qubit storage and two-way classical communication, does require atomic qubits at each repeater station for the local state preparations and the error correction operations. Therefore, it cannot simply be operated at the potentially very fast clock rate of a laser source or an optical quantum state preparation device. Instead, the elementary time units of our scheme are determined by the speed of the local atom-light operations (typically operating at MHz rates). Another limiting factor will be the small repeater spacing which is required for the bosonic quantum error correction that works best at sufficiently small channel loss. In this case, the local experimental imperfections of the atom-light system at each repeater station will accumulate more heavily. Here we model these local imperfections of the CQED system at each station as an additional constant loss. There is no time dependence of the local loss, since it is not associated with a variable memory decay in our memoryless scheme. Nonetheless, local loss is a bottle neck in our scheme, because it must be very small in order to allow for the frequent error correction steps along the channel in our CQED-based repeater scheme.

## 2. Background and Methods

It has been shown both in theory[20,21] and in experiment[5] that so called even or odd cat states (quantum superposition of optical coherent states) can be generated with the help of a high-finesse cavity containing a single trapped atom. The trapped atom consists of three relevant levels. Two of them are the ground state, $|\uparrow\rangle$ and $|\downarrow\rangle$, the remaining one is the excited state, $|e\rangle$. The transition $|\uparrow\rangle \rightarrow |e\rangle$ is strongly coupled to the bare cavity mode. The transition $|\downarrow\rangle \rightarrow |e\rangle$ is decoupled from the cavity mode because of the energy gap between the two ground states. Then one may consider the scenario that an input light field which is resonant with the cavity mode is reflected from the cavity. If the atom is prepared in state $|\uparrow\rangle$, then due to the strong coupling of the cavity and the atom, the frequency of the dressed cavity mode will be significantly detuned from that of the input light. In this case the reflection becomes similar to the reflection from a mirror, which keeps the input state of light unchanged. If the atom is prepared in state $|\downarrow\rangle$, the input light mode is strongly coupled with the cavity mode, so after the reflection, the output mode will acquire an extra phase of $e^{i\pi}$. However, if the light mode is not perfectly resonant with the cavity mode, i.e., there is a detuning $\Delta$ between them, the extra phase becomes different from $e^{i\pi}$. According to the quantum optics calculation, the relation between the input and output light mode operators is[20]

$$\hat{a}_{out} \approx \frac{i\Delta - \frac{\kappa}{2}}{i\Delta + \frac{\kappa}{2}} \hat{a}_{in}, \qquad (1)$$

where $\kappa$ is the cavity decay rate. This relation corresponds to the rotation operator

$$\hat{R}(\phi) = \exp(i\phi \hat{n}) \qquad (2)$$

acting on the state of the input light, where the rotation angle $\phi = \arg\left((i\Delta - \frac{\kappa}{2})/(i\Delta + \frac{\kappa}{2})\right)$. With this quantum gate for optical modes, we are able to create states with rotational symmetries. Moreover, for quantum error correction using certain suitably encoded states, this operation can extract information that tells us which error space a state belongs to. This kind of information is part of the syndrome measurement, which is crucial to enhance fidelities when the states are subject to errors. Finally, this provides a possibility to generate entanglement probabilistically after an appropriate measurement, for instance, via unambiguous state discrimination (**USD**) or homodyne detection.

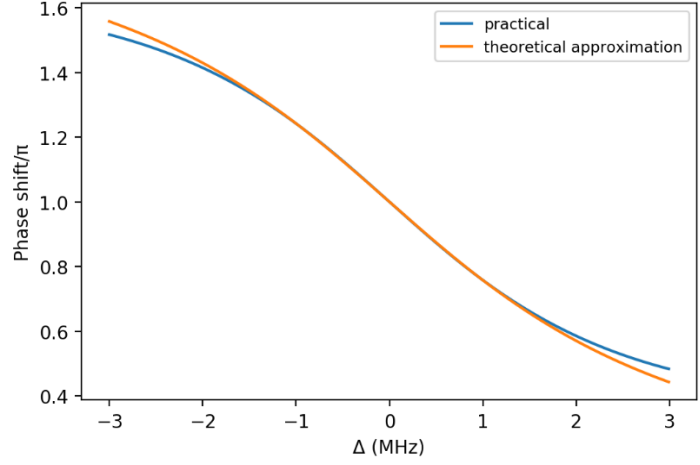

**Figure 1.** The phase shift caused by the light-matter interaction as a function of the detuning Δ. The orange curve is the theoretical approximation which keeps the amplitude unchanged and the blue one, including more experimental parameters, is more accurate, but the amplitude is reduced.

In experiments, there are more parameters related to the rotation angle $\phi$ and the amplitude will be damped slightly after the reflection because of the inevitable losses.[5] The respective complex reflection amplitude $r$ including additional parameters is,[5]

$$r(\Delta) = 1 - \frac{2\kappa_r(2i\pi\Delta + \gamma)}{(2i\pi\Delta + \kappa)(2i\pi\Delta + \gamma) + g^2}. \tag{3}$$

where $\kappa_r/\kappa$ is the reduced escape efficiency of the cavity, $\gamma$ is the atomic dipole decay rate and $g$ is the atom-cavity coupling constant. We set these parameters to be the same as in Ref. [5] in **Figure 1**. From Figure 1, one can see that with proper parameters (to be more specific, it is required that $g^2 \gg \kappa\gamma$ and $\kappa_r \approx \kappa$), the phase shift almost remains the same. At the beginning, in order to describe our repeater protocol, we will assume perfect local operations (no local loss) and later in the rate analysis part, we will also consider the local losses including the amplitude damping as represented by Equation 3. Figure 1 shows that in order to get a small phase shift close to zero, one needs a rather large detuning. The large detuning might not be easily realizable in the lab compared to a smaller detuning and a larger detuning might also affect the modulus of the amplitude of the output mode as shown in Equation 3. In this section, for simplicity, we make use of such cavities with a rotation angle $\phi$ smaller than $\pi/2$, although it may not be practical to prepare such cavities in the lab. Fortunately, all the cavities can be replaced by those with $\phi$ bigger than $\pi/2$ and smaller than $\pi$ (see S2). Thus, all the discussions in this section are still valid, as long as the discussion as presented in S2 is taken into account with regards to real experiments.

**In the next three subsections, we first consider only a single repeater segment to do the quantum error correction and finally create the entanglement. Quantum memories may be present at the station that sends the light mode (Alice). If we want to create entanglement shared between Alice and Bob and keep it for certain applications, we need another quantum memory at the station receiving the light mode (Bob). However, in our final, complete quantum repeater scheme, connecting many such elementary segments, the overall procedure can be done without quantum memories and only the entangled pairs in Alice and Bob's hands may be kept. When the specific application is QKD, Alice's atom can be measured once the desired code states are available at the station sending the light mode. Similarly, also Bob's atom can be measured immediately after all the necessary light-atom interactions at the station receiving the light mode. So, no quantum memories are needed at all (even for Alice and Bob) for QKD. Our theoretical analysis and all derivations are independent of the particular choice of application (see Section 3.1 and S7 for details). Hence it works for both long-range QKD and applications that rely on long-distance stationary entanglement distribution between Alice and Bob.**

## 2.1. State Preparation

The method mentioned above provides a realization for a kind of 'hybrid controlled rotation gate'. Here 'hybrid' means that the controlling part is the atomic state and the target part is the light mode, and so the gate performs a hybrid operation between the atomic state and the incident light mode. Similarly, the gate is expressible by discrete-variable (DV) qubit operators acting on the atom and continuous-variable (CV) oscillator ('qumode') operators acting on the light field mode. With the help of this gate, we can easily generate states with discrete rotation symmetry. First, we prepare the atom as mentioned above in a superposition of the two ground states, $(|\uparrow\rangle + |\downarrow\rangle)/\sqrt{2}$, and then reflect an arbitrary light-mode state $|\Theta\rangle$ from the cavity. Then, as mentioned before, the light mode will be either rotated in phase space or remain the same depending on the state of the atom,

$$|\uparrow\rangle|\Theta\rangle \rightarrow |\uparrow\rangle|\Theta\rangle,$$

$$|\downarrow\rangle|\Theta\rangle \rightarrow |\downarrow\rangle \hat{R}(\phi)|\Theta\rangle. \tag{4}$$

This action can be formally expressed as a hybrid CROT (controlled rotation) gate acting jointly on an atomic and a light mode input state,

$$h\text{CROT}_\phi = |\uparrow\rangle\langle\uparrow|\otimes \mathbb{1} + |\downarrow\rangle\langle\downarrow|\otimes \exp(i\phi\hat{n}). \tag{5}$$

To create a state of light with $2^m$-fold rotation symmetry, at first we need to set the rotation angle $\phi = \pi$, which can be realized by setting the detuning $\Delta = 0$. Thus, after the first reflection, the state becomes

$$\frac{1}{\sqrt{2}}\big(|\uparrow\rangle|\Theta\rangle + |\downarrow\rangle\hat{R}(\pi)|\Theta\rangle\big), \tag{6}$$

so the atom and the light mode become entangled. We can then rewrite the state,

$$\frac{1}{\sqrt{2}}\big(|\uparrow\rangle|\Theta\rangle + |\downarrow\rangle\hat{R}(\pi)|\Theta\rangle\big)$$
$$= \frac{1}{2}\left[\frac{1}{\sqrt{2}}(|\uparrow\rangle + |\downarrow\rangle)\big(|\Theta\rangle + \hat{R}(\pi)|\Theta\rangle\big) + \frac{1}{\sqrt{2}}(|\uparrow\rangle - |\downarrow\rangle)\big(|\Theta\rangle - \hat{R}(\pi)|\Theta\rangle\big)\right]. \tag{7}$$

Next, after doing a projective measurement of the atomic state in the basis $\{|\pm\rangle = (|\uparrow\rangle \pm |\downarrow\rangle)/\sqrt{2}\}$, the state of light will be projected onto a superposition of $|\Theta\rangle$ and the rotated state $\hat{R}(\pi)|\Theta\rangle$, i.e., $|\Theta\rangle \pm \hat{R}(\pi)|\Theta\rangle$ (unnormalized). This step has already been realized in the lab with $|\Theta\rangle = |\alpha\rangle$, generating the even or odd 'cat states' $(|\alpha\rangle \pm |-\alpha\rangle)/N_\pm$.[5] As a next step, we adjust the cavity (or prepare another cavity) to make the detuning $\Delta = \kappa/2$, so the rotation angle $\phi$ becomes $\pi/2$, and we prepare the atom inside again in the state $(|\uparrow\rangle + |\downarrow\rangle)/\sqrt{2}$.

Assuming we obtain $|\Theta\rangle + \hat{R}(\pi)|\Theta\rangle$ from the first step, then if we reflect this state again from the adjusted cavity (or another cavity), we will get the light state

$$|\Theta\rangle + \hat{R}(\pi)|\Theta\rangle \pm \hat{R}\left(\frac{\pi}{2}\right)|\Theta\rangle \pm \hat{R}\left(\frac{3\pi}{2}\right)|\Theta\rangle \tag{8}$$

after the projective measurement of the atomic state similar to the first step.[22] Repeating the steps above and adjusting the cavity (or preparing another cavity) in every step with the rotation angle $\phi$ being half of the angle from the previous step, finally, the joint atom-light state (before the atomic measurement) will become

$$\frac{1}{\sqrt{2}}\left(|\uparrow\rangle|0_{2^m,\Theta}\rangle + |\downarrow\rangle|1_{2^m,\Theta}\rangle\right), \tag{9}$$

$\forall m \in \mathbb{N}_0$ after $m + 1$ steps, where

$$|0_{M,\Theta}\rangle = \frac{1}{\sqrt{N_M}} \sum_{k=0}^{M-1} \hat{R}\left(\frac{2k\pi}{M}\right)|\Theta\rangle,$$

$$|1_{M,\Theta}\rangle = \frac{1}{\sqrt{N_M}} \sum_{k=0}^{M-1} \hat{R}\left(\frac{(2k+1)\pi}{M}\right)|\Theta\rangle, \tag{10}$$

with a normalization constant $N_M$. The two states $|0_{M,\Theta}\rangle$ and $|1_{M,\Theta}\rangle$ are the logical codewords used in our scheme. Both of them exhibit rotation symmetry, i.e.,

$$\hat{R}\left(\frac{2\pi}{M}\right)|0_{M,\Theta}\rangle = |0_{M,\Theta}\rangle,$$

$$\hat{R}\left(\frac{2\pi}{M}\right)|1_{M,\Theta}\rangle = |1_{M,\Theta}\rangle. \tag{11}$$

Equation 10 is the general definition of the two codewords of a rotation-symmetric bosonic qubit code for any integer $M$. However, according to Equation 9, the above method can only generate states for $M = 2^m$.

Here the codewords are generally not orthogonal, $\langle 0_{M,\Theta}|1_{M,\Theta}\rangle \neq 0$. Nevertheless, the logical codewords can also be constructed to be orthogonal.[8] For example, the two orthogonal states $|\Theta\rangle + \hat{R}(\pi)|\Theta\rangle$ and $|\Theta\rangle - \hat{R}(\pi)|\Theta\rangle$ (unnormalized) in Equation 7 can also be used as the codewords. Although the orthogonal codewords can be discriminated deterministically (at least in theory), they cannot lead to a deterministic entanglement creation process, because one has to add two different normalization constants to normalize the two states and the difference will finally make the process probabilistic again. A more detailed discussion can be found in S3.

**2.2. Syndrome Measurement**

Quantum error correction (QEC) is essential to suppress the errors, in our case, primarily related with the LBC. In order to correct the errors, a syndrome measurement is necessary, which extracts the error information from the encoded state.

After preparing the state in Equation 9 in the desired loss order (the loss order $L = 2^m - 1$ for $|0_{M,\Theta}\rangle$ and $|1_{M,\Theta}\rangle$.), 'Alice' transmits the light mode via an optical fiber. Inevitably, some photons will get lost during the transmission. Before considering the full loss channel, it is conceptually useful to first think of a simplified loss channel, where we only apply the annihilation operator $\hat{a}$ to the codewords. Higher losses then correspond to higher powers of $\hat{a}$. The resulting state becomes

$$\frac{1}{\sqrt{2}}\left(|\uparrow\rangle_A \hat{a}^q |0_{2^m,\Theta}\rangle + |\downarrow\rangle_A \hat{a}^q |1_{2^m,\Theta}\rangle\right) \tag{12}$$

if $q$ photons are lost. Here $|\uparrow\rangle_A$ and $|\downarrow\rangle_A$ represent the atomic state held by Alice. Analogously to the generation of this state, we also do the syndrome measurement with the help of a cavity. We adjust the cavity with the rotation angle $\phi = \pi/2^{m-1}$ and the state of the atom inside is prepared to be $\left(|\uparrow\rangle_s + |\downarrow\rangle_s\right)/\sqrt{2}$. When we reflect the light mode after the transmission from this cavity, the joint atom-light state becomes

$$\frac{1}{2}|\uparrow\rangle_s \left(|\uparrow\rangle_A \hat{a}^q |0_{2^m,\Theta}\rangle + |\downarrow\rangle_A \hat{a}^q |1_{2^m,\Theta}\rangle\right)$$
$$+ \frac{1}{2}|\downarrow\rangle_s \left(|\uparrow\rangle_A \hat{R}\left(\frac{\pi}{2^{m-1}}\right)\hat{a}^q |0_{2^m,\Theta}\rangle + |\downarrow\rangle_A \hat{R}\left(\frac{\pi}{2^{m-1}}\right)\hat{a}^q |1_{2^m,\Theta}\rangle\right), \tag{13}$$

where $|\uparrow\rangle_s$ and $|\downarrow\rangle_s$ represent the state of the ancillary atom used for the syndrome measurement. There is a relation of the rotation operator $\hat{R}(\phi)$ and the annihilation operator $\hat{a}$,

$$\begin{aligned}\hat{R}(\phi)\hat{a}^q &= \hat{R}(\phi)\hat{a}^q\hat{R}^\dagger(\phi)\hat{R}(\phi) = \left(\hat{R}(\phi)\hat{a}\hat{R}^\dagger(\phi)\right)^q \hat{R}(\phi) \\ &= \left(e^{-i\phi}\hat{a}\right)^q \hat{R}(\phi) = e^{-iq\phi}\,\hat{a}^q\hat{R}(\phi).\end{aligned} \quad (14)$$

Also, according to Equation 11, we have

$$\hat{R}\left(\frac{\pi}{2^{m-1}}\right)|0_{2^m,\Theta}\rangle = |0_{2^m,\Theta}\rangle,$$

$$\hat{R}\left(\frac{\pi}{2^{m-1}}\right)|1_{2^m,\Theta}\rangle = |1_{2^m,\Theta}\rangle. \quad (15)$$

Then Equation 13 can be rewritten as

$$\frac{1}{2}\left(|\uparrow\rangle_s + e^{-\frac{iq\pi}{2^{m-1}}}|\downarrow\rangle_s\right)\left(|\uparrow\rangle_A\hat{a}^q|0_{2^m,\Theta}\rangle + |\downarrow\rangle_A\hat{a}^q|1_{2^m,\Theta}\rangle\right), \quad (16)$$

which is a product of the atomic state $\left(|\uparrow\rangle_s + \exp\left(-iq\pi/2^{m-1}\right)|\downarrow\rangle_s\right)/\sqrt{2}$ and the state in Equation 12. Thus, with the help of this kind of cavity, we can construct a quantum nondemolition (QND) measurement, which contains information in the atomic state of the syndrome spin about how many photons are lost from the light mode after transmission.

Unfortunately, there are also some constraints in such a scheme. First, the extra phase obtained in the atomic state $\exp\left(-iq\pi/2^{m-1}\right)$ is not unique for every $q$ with finite $m$, e.g., for any $q_1 \in \mathbb{N}_0$, if we choose $q_2 = q_1 + 2^m$, there is a relation $\exp\left(-iq_2\pi/2^{m-1}\right) = \exp\left(-iq_1\pi/2^{m-1}\right)$. This means there can be states with uncorrectable errors which cannot be distinguished by this method. This effect will decrease the total fidelity depending on the choice of the primitive state $|\Theta\rangle$. In Section 3, we will analyze in detail the performance of our repeater scheme for the special case of 'cat codes'. Secondly, it is not always possible to deterministically discriminate the atomic states with different phases, even in theory, because in general they are not mutually orthogonal.[23,24] However, this complication can be circumvented by employing more cavities (or adjusting the same cavity many times). The idea is to extract partial information of $q$ at several times and then combine all the partial information to get the full syndrome information. For this purpose, we first reflect the light mode in Equation 12 from a cavity with the rotation angle $\phi = \pi$. The resulting state is

$$\frac{1}{2}\left(|\uparrow\rangle_s + e^{-iq\pi}|\downarrow\rangle_s\right)\left(|\uparrow\rangle_A \hat{a}^q |0_{2^m,\Theta}\rangle + |\downarrow\rangle_A \hat{a}^q |1_{2^m,\Theta}\rangle\right), \tag{17}$$

where the extra phase $e^{-iq\pi} = 1$ if $q$ is even and $e^{-iq\pi} = -1$ if $q$ is odd. It is not difficult to discriminate the two orthogonal atomic states $|\uparrow\rangle_s + |\downarrow\rangle_s$ and $|\uparrow\rangle_s - |\downarrow\rangle_s$ and after the measurement of the atomic state, we have some information of $q$, i.e., whether it is even or odd. Next, we repeat this process but prepare the cavity with the rotation angle $\phi = \pi/2$ and reflect the light mode again. Not surprisingly, this time the output state becomes,[25]

$$\frac{1}{2}\left(|\uparrow\rangle_s + e^{-\frac{iq\pi}{2}}|\downarrow\rangle_s\right)\left(|\uparrow\rangle_A \hat{a}^q |0_{2^m,\Theta}\rangle + |\downarrow\rangle_A \hat{a}^q |1_{2^m,\Theta}\rangle\right). \tag{18}$$

Now we look at the cases when $q$ is even or odd separately. In the case when $q$ is even, $e^{-iq\pi/2} = 1$ if $q \bmod 4 = 0$, $e^{-iq\pi/2} = -1$ if $q \bmod 4 = 2$. For the case that $q$ is odd, $e^{-iq\pi/2} = -i$ if $q \bmod 4 = 1$, $e^{-iq\pi/2} = i$ if $q \bmod 4 = 3$. In either case, the atomic states we need to discriminate are orthogonal (even: Pauli $X$ eigenstates, odd: Pauli $Y$ eigenstates). So in either case, the atomic states can be discriminated deterministically, at least in principle, by adapting the spin basis of the second measurement according to the outcome of the first (or by including a corresponding spin rotation prior to the second measurement depending on the result of the first). The steps above can be repeated again and again if we divide the rotation angle $\phi$ in half at every step until $\phi = \pi/2^{m-1}$ (see S1). After all the steps, we will finally know the remainder of $q$ divided by $2^m$, $r_m(q)$, i.e., $q \bmod 2^m = r_m(q)$. The $r_m(q)$ tells us which error space (or the code space) the light mode lies in and this is crucial to improve the total fidelity.

## 2.3. Entanglement Creation

In our protocol, the transmitted light mode is used to create the entanglement for the atomic states that Alice and Bob hold and the cavity-assisted method can also be used to create such entanglement. In order to do this, Bob needs to prepare another such cavity after his syndrome measurement and this time the rotation angle $\phi$ should be $\pi/2^m$. With this angle, a 'bit flip' happens after the corresponding rotation operator $\hat{R}(\phi)$ acts on the two codewords in Equation 9, i.e.,

$$\hat{R}\left(\frac{\pi}{2^m}\right)|0_{2^m,\Theta}\rangle = |1_{2^m,\Theta}\rangle,$$

$$\hat{R}\left(\frac{\pi}{2^m}\right)|1_{2^m,\Theta}\rangle = |0_{2^m,\Theta}\rangle. \quad (19)$$

This relation can be easily derived from Equation 10. Thus, the state after this interaction becomes (using again Equation 14 and now, in addition, Equation 19)

$$\frac{1}{2}|\uparrow\rangle_B\left(|\uparrow\rangle_A \hat{a}^q|0_{2^m,\Theta}\rangle + |\downarrow\rangle_A \hat{a}^q|1_{2^m,\Theta}\rangle\right)$$
$$+\frac{1}{2}e^{-\frac{iq\pi}{2^m}}|\downarrow\rangle_B\left(|\uparrow\rangle_A \hat{a}^q|1_{2^m,\Theta}\rangle + |\downarrow\rangle_A \hat{a}^q|0_{2^m,\Theta}\rangle\right)$$
$$=\frac{1}{2}\left[\left(|\uparrow\rangle_B|\uparrow\rangle_A + e^{-\frac{iq\pi}{2^m}}|\downarrow\rangle_B|\downarrow\rangle_A\right)\hat{a}^q|0_{2^m,\Theta}\rangle + \left(|\uparrow\rangle_B|\downarrow\rangle_A + e^{-\frac{iq\pi}{2^m}}|\downarrow\rangle_B|\uparrow\rangle_A\right)\hat{a}^q|1_{2^m,\Theta}\rangle\right]. \quad (20)$$

Because this interaction happens after the syndrome measurement, $r_m(q)$ is supposed to be known. Finally, according to Equation 20, the discrimination between the light states $\hat{a}^q|0_{2^m,\Theta}\rangle$ and $\hat{a}^q|1_{2^m,\Theta}\rangle$ (actually, the states to be discriminated are not exactly these two states if the full loss channel is considered, as we shall discuss in Section 3.4 in more detail) leads to a collapse for the state in Equation 20 and the conditional atomic state of Alice and Bob will always become entangled. However, $\hat{a}^q|0_{2^m,\Theta}\rangle$ and $\hat{a}^q|1_{2^m,\Theta}\rangle$ are generally not orthogonal, thus the discrimination cannot be done deterministically (or otherwise in an error-free fashion). This is the only nondeterministic element in our scheme if we choose an error-free, unambiguous discrimination. There are several ways to do the state discrimination, with distinct advantages and drawbacks. A detailed discussion will be presented in Section 3. The final entangled state will be either $|\uparrow\rangle_B|\uparrow\rangle_A + \exp(-iq\pi/2^m)|\downarrow\rangle_B|\downarrow\rangle_A$ or $|\uparrow\rangle_B|\downarrow\rangle_A + \exp(-iq\pi/2^m)|\downarrow\rangle_B|\uparrow\rangle_A$, depending on the measurement result for the light mode. Again note that $q$ is assumed to be known, so we have full information of the final entangled state, which means this kind of entanglement can be a resource for quantum communication.

**2.4. Full Loss Channel**

All the discussions in Sections 2.2 and 2.3 are based on the simplified loss channel, but in practice, photon loss is described by the full, physical amplitude damping (AD) channel.[6,26] Generally, a single-mode state after the action of AD is mixed and it can be expressed as

$$\hat{\rho}' = \sum_{k=0}^{\infty} \hat{A}_k \hat{\rho} \hat{A}_k^\dagger, \tag{21}$$

where $\hat{\rho}$ and $\hat{\rho}'$ are the density operators of the states before and after the action of the channel, respectively. Here, $\hat{A}_k$ is a nonunitary error operator,[6,26]

$$\hat{A}_k = \sum_{n=k}^{\infty} \sqrt{\binom{n}{k}} \sqrt{\eta^{n-k}(1-\eta)^k} |n-k\rangle\langle n| = \sqrt{\frac{(1-\eta)^k}{k!}} \sqrt{\eta}^{\hat{n}} \hat{a}^k, \quad \forall k \in \mathbb{N}_0, \tag{22}$$

where $\eta$ is the transmission of the fiber, i.e., the probability of losing one photon is $1-\eta$.

Then the action of AD on the state in Equation 9 leads to a mixed state and the final density operator becomes

$$\hat{\rho}_f = \sum_{k=0}^{\infty} \hat{A}_k \frac{1}{\sqrt{2}} \left(|\uparrow\rangle|0_{2^m,\Theta}\rangle + |\downarrow\rangle|1_{2^m,\Theta}\rangle\right) \times \text{H.c.} \tag{23}$$

Next, if there is a rotation operator acting on it, the density operator transforms as

$$\hat{R}(\phi)\hat{\rho}_f \hat{R}^\dagger(\phi) = \sum_{k=0}^{\infty} \hat{R}(\phi)\hat{A}_k \frac{1}{\sqrt{2}} \left(|\uparrow\rangle|0_{2^m,\Theta}\rangle + |\downarrow\rangle|1_{2^m,\Theta}\rangle\right) \times \text{H.c.} \tag{24}$$

If we only look at the $k$-component of Equation 24 and temporarily ignore the Hermitian conjugate of it, i.e.,

$$\hat{R}(\phi)\hat{A}_k \frac{1}{\sqrt{2}}\left(|\uparrow\rangle|0_{2^m,\Theta}\rangle + |\downarrow\rangle|1_{2^m,\Theta}\rangle\right)$$

$$= e^{i\phi\hat{n}} \sqrt{\frac{(1-\eta)^k}{k!}} \sqrt{\eta}^{\hat{n}} \hat{a}^k \frac{1}{\sqrt{2}}\left(|\uparrow\rangle|0_{2^m,\Theta}\rangle + |\downarrow\rangle|1_{2^m,\Theta}\rangle\right)$$

$$= \left(\sqrt{\frac{(1-\eta)^k}{k!}} \sqrt{\eta}^{\hat{n}}\right) e^{i\phi\hat{n}} \hat{a}^k \frac{1}{\sqrt{2}}\left(|\uparrow\rangle|0_{2^m,\Theta}\rangle + |\downarrow\rangle|1_{2^m,\Theta}\rangle\right).$$

$$= e^{-ik\phi} \left(\sqrt{\frac{(1-\eta)^k}{k!}} \sqrt{\eta}^{\hat{n}} \hat{a}^k\right) e^{i\phi\hat{n}} \frac{1}{\sqrt{2}}\left(|\uparrow\rangle|0_{2^m,\Theta}\rangle + |\downarrow\rangle|1_{2^m,\Theta}\rangle\right)$$

$$= e^{-ik\phi} \frac{1}{\sqrt{2}}\left(|\uparrow\rangle\hat{A}_k\hat{R}(\phi)|0_{2^m,\Theta}\rangle + |\downarrow\rangle\hat{A}_k\hat{R}(\phi)|1_{2^m,\Theta}\rangle\right). \tag{25}$$

Comparing Equation 25 with Equation 16, including the steps for the syndrome detection in Equations 13-16, it is obvious that for the $k$- or $q$-components, the only difference is that $\hat{a}^k$ is replaced by $\hat{A}_k$, bringing a prefactor $\hat{A}_k/\hat{a}^k = \sqrt{(1-\eta)^k/k!}\sqrt{\eta}^{\hat{n}}$. Thus, all the results in Sections 2.2 and 2.3 are exactly applicable for the full loss channel up to these prefactors acting on each component. It is not straightforward what effects these prefactors will bring for an arbitrary primitive state $|\Theta\rangle$. As an example of our scheme, we refer to a part of the results in Ref. [6] and evaluate the performance of our repeater protocol using the 'cat codes'.

## 3. Analysis

In the preceding section, we described the basic elements of our protocol that could be used, in principle, for quantum communication over long distances at high rates. Here in this section, we explicitly propose such a scheme for long-distance quantum communication based on these elements. We will evaluate the performance of this scheme numerically for a particular case of the rotation-symmetric bosonic codes—the cat codes.

## 3.1. Scheme for Long-Distance Quantum Communication

We want to propose a scheme without the need of quantum memories, similar to a third-generation quantum repeater,[3] with the goal not to slow down the repeater due to two-way classical communication and the corresponding waiting times. But instead of directly transmitting encoded qubits, in our scheme, we aim to create entangled states shared between Alice and Bob over a long distance.

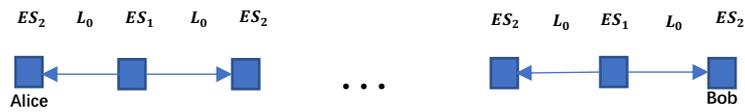

**Figure 2.** Schematic of the long-distance entanglement distribution protocol including two types of elementary stations: $ES_1$ to send the light mode and $ES_2$ to receive it.

Realizing long-distance quantum communication directly from one point (Alice) to another (Bob) connected by a fiber channel is not efficient because of the photon loss in the fiber, and so intermediate stations have to be employed along the channel between Alice and Bob in order to suppress the exponential decay of either qubit rates or qubit state fidelities. Assume that the total distance between Alice and Bob is $L_{tot}$, then the repeater stations are placed at every elementary distance $L_0 = L_{tot}/n_e$, where $n_e$ is the number of elementary links. The repeater stations are divided into two different types according to their function: one is for sending the light and the other one is for receiving it (see Figure 2).

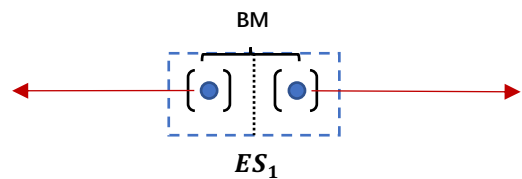

(a) Structure of $ES_1$

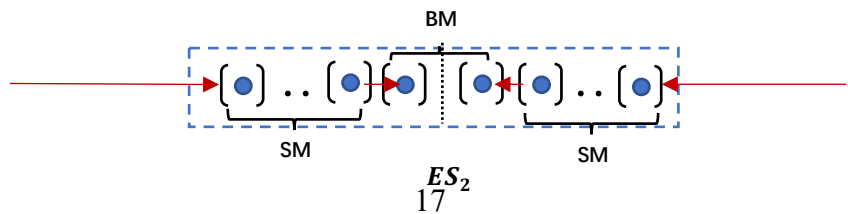

$ES_2$

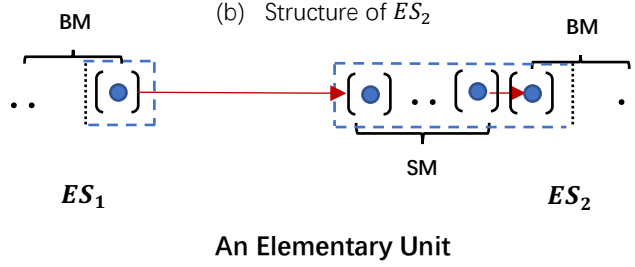

(c) Structure of an elementary 'unit'

**Figure 3.** The structure of two types of elementary stations: (a) $ES_1$ and (b) $ES_2$, (c) one half of $ES_1$ and one half of $ES_2$ together can form an elementary 'unit'. The different spin measurements are denoted by BM (Bell measurement) and SM (syndrome measurement).

The structure of the two kinds of elementary stations is depicted in **Figure 3**. $ES_1$, which sends the light, consists of two cavities. The cavities are used to generate the desired light mode state according to Section 2.1 and the two modes just generated are then sent to the two nearest stations separately in opposite directions. For $ES_2$, which receives the light, the number of involved cavities varies depending on the loss order of the code. The syndrome measurements and the entanglement creation are performed in $ES_2$ as described in Sections 2.2 and 2.3. Note that the number of cavities needed in each repeater station can be different from what is depicted in Figure 3, because for the generation and the syndrome measurement procedures, one can either use one cavity many times with different parameters or use many different cavities only once. It is obvious that $ES_1$ and $ES_2$ are mirror-symmetric, so both can be divided into two parts. The right part of $ES_1$ and the left part of $ES_2$ (or the other way around) can form a basic elementary unit of the whole protocol. In an elementary unit, all the three procedures including state preparation, syndrome measurement, and entanglement creation are performed in corresponding order. This results in an entangled state between the atomic spin in $ES_1$ and the atom inside the last cavity that interacted with the light mode in $ES_2$. Thus, many such elementary units connecting each other can form the whole repeater protocol. Bell measurements are then performed on the two still unmeasured spins at every repeater station to swap the entanglement and hence coherently connect the individual segments. Finally, the atomic spins in Alice's and Bob's hands are entangled over the total distance (at least 'effectively' entangled in case Alice and Bob decide to measure their qubits immediately depending on the repeater's application).

In practice, the Bell measurement on the two atoms at every $ES_1$ can be done directly after the state preparation process. This way the entanglement between the atoms in every half of $ES_1$ and the corresponding light mode sent away becomes the entanglement between the two light modes that $ES_1$ has sent. So, there is no need to use quantum memories at $ES_1$ to store the -light entanglement (but memories may still be employed by Alice and Bob if they want to keep their entangled pairs). The sender and receiver stations (of Alice and Bob) do not constitute a complete $ES_1$ or $ES_2$, but only a half of it. At each $ES_1$ which connects two nearest 'units', the state preparation at the two parts must finish at the same time and the Bell measurement will be done immediately after the state preparation (so now the two flying light modes get entangled). We assume that all the elementary distances between two nearest stations are the same. Then the light modes will arrive at every $ES_2$ simultaneously, followed by the syndrome measurement and the entanglement creation. After all the processes, the atoms at the $ES_2$ parts in the nearest 'units' connected by $ES_1$ get entangled. Finally, after doing Bell measurements at all the $ES_2$'s, the atomic states in Alice's and Bob's hands are entangled. More details of the derivations can be found in S7. Also, if we do not need to keep the entangled pairs for Alice and Bob, e.g., when doing QKD, Alice and Bob can measure their atoms immediately and do not need the memories either. In conclusion, as long as all the events are sufficiently well synchronized, no quantum memories are needed in the whole scheme to realize long-distance quantum communication. The precise moment when we do the Bell measurements at the $ES_1$'s is irrelevant to the final joint state for Alice and Bob, so for simplicity and without loss of generality, the related theoretical derivations are mostly based on what happens in a single elementary 'unit'.

For every elementary unit, there is a non-unit fidelity $F_0$ for the distributed spin-spin entangled states and a non-unit success probability $P_0$ for the state discrimination used to create the entanglement. The total success probability $P_{tot}$ is the probability that the state discrimination succeeds at every station, so it is easy to get the relation, $P_{tot} = P_0^{n_e}$. To calculate the total fidelity $F_{tot}$, it is necessary to consider the entanglement swapping process (also see S4).[27,28]

**3.2. Cat Codes**

As described in Section 2.1, for any optical states, it is possible to construct a so-called RSBC as in Equation 10. The 'cat codes' are a special case for the RSBC, for which we choose the primitive state $|\Theta\rangle$ to be the optical coherent state $|\alpha\rangle$.[4,6,29] Thus, the corresponding codewords are,[6]

$$|0_{M,cat}\rangle = \frac{1}{\sqrt{N_M}} \sum_{k=0}^{M-1} \hat{R}\left(\frac{2k\pi}{M}\right)|\alpha\rangle = \frac{1}{\sqrt{N_M}} \sum_{k=0}^{M-1} \left|\alpha e^{\frac{2k\pi i}{M}}\right\rangle,$$

$$|1_{M,cat}\rangle = \frac{1}{\sqrt{N_M}} \sum_{k=0}^{M-1} \hat{R}\left(\frac{(2k+1)\pi}{M}\right)|\alpha\rangle = \frac{1}{\sqrt{N_M}} \sum_{k=0}^{M-1} \left|\alpha e^{\frac{(2k+1)\pi i}{M}}\right\rangle. \quad (26)$$

It is then straightforward to find out the cyclic behavior for cat codes in the simplified loss model,[6]

$$\hat{a}^{Mn}|0_{M,cat}\rangle = \alpha^{Mn}|0_{M,cat}\rangle,$$

$$\hat{a}^{Mn}|1_{M,cat}\rangle = (-1)^n \alpha^{Mn}|1_{M,cat}\rangle, \quad (27)$$

$\forall n \in \mathbb{N}_0$. The cyclic rule means that for cat codes, the codewords or the states in an error space will come back to the corresponding code or error space, which contributes to the total fidelity. Recall that the method in Section 2.1 can only generate states with $M = 2^m$.

After some lengthy calculations,[6] the density matrix of the joint state of the atom (at Alice again considering first a single link between Alice and Bob) and the light mode after the LBC becomes

$$\hat{\rho}' = \sum_{q=0}^{2^{m+1}-1} p_q \left( \frac{|\uparrow\rangle_A \hat{a}^q |\tilde{0}_{2^m,cat}\rangle + |\downarrow\rangle_A \hat{a}^q |\tilde{1}_{2^m,cat}\rangle}{\sqrt{2N'_{m,q}}} \right) \times \text{H.c.}, \tag{28}$$

where the statistical weights are (assuming real $\alpha$)

$$p_q = \frac{1}{\sqrt{N'_M}} \sum_{j=0}^{\infty} \frac{[\alpha^2(1-\eta)]^{2^{m+1}\cdot j+q}}{(2^{m+1}\cdot j+q)!}. \tag{29}$$

Here, $|\tilde{0}_{2^m,cat}\rangle$ and $|\tilde{1}_{2^m,cat}\rangle$ are the damped codewords,

$$|\tilde{0}_{2^m,cat}\rangle = \frac{1}{\sqrt{N'_M}} \sum_{k=0}^{M-1} \hat{R}\left(\frac{2k\pi}{M}\right) |\sqrt{\eta}\alpha\rangle,$$

$$|\tilde{1}_{2^m,cat}\rangle = \frac{1}{\sqrt{N'_M}} \sum_{k=0}^{M-1} \hat{R}\left(\frac{(2k+1)\pi}{M}\right) |\sqrt{\eta}\alpha\rangle, \tag{30}$$

which means the amplitudes of the coherent-state components are damped. The states in loss space — $\hat{a}^q |\tilde{0}_{2^m,cat}\rangle$ and $\hat{a}^q |\tilde{1}_{2^m,cat}\rangle$ — are not normalized anymore, so a normalization constant $N'_{m,q}$ is needed to normalize them ($\hat{a}^q |\tilde{0}_{2^m,cat}\rangle / \sqrt{N'_{m,q}}$ and $\hat{a}^q |\tilde{1}_{2^m,cat}\rangle / \sqrt{N'_{m,q}}$ are normalized states now). Because of the cyclic behavior presented in Equation 27, all the higher terms with $q \geq 2^{m+1}$ have their corresponding terms with $q < 2^{m+1}$ and they can be simply added up. The terms of Equation 28 can be divided into two parts—the first $2^m$ terms and the last $2^m$ terms. According to Section 2.2, the first $2^m$ terms can be individually distinguished in a deterministic fashion and the last $2^m$ terms are uncorrectable errors. So, after the syndrome measurement, the density matrix will collapse to

$$\hat{\rho}'' = \frac{p_q}{p_q + p_{q+2^m}} \left( \frac{|\uparrow\rangle_A \hat{a}^q |\tilde{0}_{2^m,cat}\rangle + |\downarrow\rangle_A \hat{a}^q |\tilde{1}_{2^m,cat}\rangle}{\sqrt{2N'_{M,q}}} \right) \times \text{H.c.}$$

$$+\frac{p_{q+2^m}}{p_q+p_{q+2^m}}\left(\frac{|\uparrow\rangle_A \hat{a}^q |\tilde{0}_{2^m,cat}\rangle - |\downarrow\rangle_A \hat{a}^q |\tilde{1}_{2^m,cat}\rangle}{\sqrt{2N'_{M,q}}}\right) \times \text{H.c.}, \quad (31)$$

with probability $p_q + p_{q+2^m}$, which depends on the result of the measurement of the atomic state (here $q < 2^m$ and we make use of the cyclic behavior). According to Section 2.3, after another cavity reflection, the state becomes

$$\hat{\rho}''' = \frac{p_q}{p_q+p_{q+2^m}}\left(\frac{|\Phi^+_{q\pi/2^m}\rangle_{BA} \hat{a}^q |\tilde{0}_{2^m,cat}\rangle + |\Psi^+_{q\pi/2^m}\rangle_{BA} \hat{a}^q |\tilde{1}_{2^m,cat}\rangle}{\sqrt{2N'_{M,q}}}\right) \times \text{H.c.}$$

$$+\frac{p_{q+2^m}}{p_q+p_{q+2^m}}\left(\frac{|\Phi^-_{q\pi/2^m}\rangle_{BA} \hat{a}^q |\tilde{0}_{2^m,cat}\rangle + |\Psi^-_{q\pi/2^m}\rangle_{BA} \hat{a}^q |\tilde{1}_{2^m,cat}\rangle}{\sqrt{2N'_{M,q}}}\right) \times \text{H.c.}, \quad (32)$$

where $|\Phi^\pm_{q\pi/2^m}\rangle_{BA}$ and $|\Psi^\pm_{q\pi/2^m}\rangle_{BA}$ are the generalizations of the four Bell states. These are defined as

$$|\Phi^\pm_\theta\rangle_{BA} = \frac{|\uparrow\rangle_B|\uparrow\rangle_A \pm e^{-i\theta}|\downarrow\rangle_B|\downarrow\rangle_A}{\sqrt{2}},$$

$$|\Psi^\pm_\theta\rangle_{BA} = \frac{|\uparrow\rangle_B|\downarrow\rangle_A \pm e^{-i\theta}|\downarrow\rangle_B|\uparrow\rangle_A}{\sqrt{2}}. \quad (33)$$

Then after the state discrimination for the light mode, the final atomic state shared by Alice and Bob (i.e. here, in a single repeater segment) becomes a mixture of $|\Phi^+_{q\pi/2^m}\rangle_{BA}$ and $|\Phi^-_{q\pi/2^m}\rangle_{BA}$ or a mixture of $|\Psi^+_{q\pi/2^m}\rangle_{BA}$ and $|\Psi^-_{q\pi/2^m}\rangle_{BA}$ depending on the results of the state discrimination. In this case, our target is the state $|\Phi^+_{q\pi/2^m}\rangle_{BA}$ or $|\Psi^+_{q\pi/2^m}\rangle_{BA}$. Because $|\Phi^+_{q\pi/2^m}\rangle_{BA}$ (or $|\Psi^+_{q\pi/2^m}\rangle_{BA}$) and $|\Phi^-_{q\pi/2^m}\rangle_{BA}$ (or $|\Psi^-_{q\pi/2^m}\rangle_{BA}$) are orthogonal, the $(q+2^m)$-terms have no contribution to the total fidelity. Finally, we obtain the fidelity in an elementary, entangled-state distribution unit (i.e., the initial fidelity in a loss-corrected repeater segment)

$$F_0 = \sum_{q=0}^{2^m-1} p_q, \tag{34}$$

which is the sum of the statistical weights of the first $2^m$ terms of $\hat{\rho}'$.

Typically, there are two choices for the discrimination of light-mode quantum states, homodyne measurement or USD based on photon measurements, and these have distinct properties. The case with homodyne measurement can be deterministic or probabilistic depending on how one defines a successful measurement outcome, but the fidelity will typically decrease with growing success probability. The USD does not affect the fidelity (i.e., it is error-free), but it cannot be done deterministically for nonorthogonal states. Also, for larger loss orders ($L > 1$), it is subtle to define a successful measurement in a suitable manner. In our model, USD seems more appropriate, even though it cannot be done deterministically. The maximal success probability for USD is,[23,24]

$$\max P_{USD} = 1 - |\langle \psi_1 | \psi_2 \rangle|, \tag{35}$$

where $|\psi_1\rangle$ and $|\psi_2\rangle$ are the states to be distinguished with equal a priori probabilities. In our case, the maximal success probability to do USD in an elementary unit is then

$$\max P_0 = 1 - |\langle \tilde{0}^q_{2^m,\Theta} | \tilde{1}^q_{2^m,\Theta} \rangle|. \tag{36}$$

Since in our case, $|\tilde{0}^q_{2^m,\Theta}\rangle$ and $|\tilde{1}^q_{2^m,\Theta}\rangle$ are not orthogonal, the maximal success probability is certainly below one for finite $\alpha$. However, provided $\alpha$ is sufficiently large, the overlap will almost vanish and then the probability can be close to unity. In this case though, with growing $\alpha$, the photon loss (hence, in Equation 32, the uncorrectable phase-flip) probability increases.

### 3.3. Numerical Analysis Based on Cat Codes

According to Section 3.2, it is obvious that the fidelity and the success probability in an elementary unit, $F_0$ and $P_0$, depend on the elementary distance $L_0$. Generally, to get near-unit $F_0$ and $P_0$, $L_0$ needs to be rather small, but a small $L_0$ leads to a big $n_e$ (at a given total distance), which occurs in the powers for the total fidelity and success probability, and eventually again decreases them. So, there is a trade-off concerning the choice of the elementary distance, but this is closely related to the choice of how to encode the information. Fortunately, in our setting, this relation is rather simple—both the total fidelity and the total success probability improve with decreasing $L_0$ (see **Figure 4**). However, in practice, having too many stations means a high experimental cost, which is undesirable. Moreover, the local experimental imperfections, especially cavity loss and more generally any form of coupling losses (see below), contribute to a greater extent for an increasing number of stations. Thus, the stations need to be as few as possible and, at the same time, $F_{tot}$ and $P_{tot}$ should be near unity.

For fixed $L_0$ and $L_{tot}$, $F_{tot}$ and $P_{tot}$ also depend on the amplitude $\alpha$ in the two codewords. Here is again a trade-off—for too large $\alpha$, photons are more likely to get lost (with increased weights of the uncorrectable loss terms) and hence $F_{tot}$ decreases, while the two codewords become more orthogonal and hence the success probability for the state discrimination increases. Conversely, for smaller $\alpha$, $F_{tot}$ becomes larger and $P_{tot}$ becomes smaller. A larger loss order $L = 2^m - 1$ is also beneficial for the fidelity, but it leads to a decreasing $P_{tot}$. As shown in Figure 4, for repeaterless cases ($L_0 = L_{tot} = 1000 km$), the fidelities do not always go down with increasing α. This can be explained by looking at what happens when α increases from zero. For example, for the 1-loss code ($L = 1$), when α is very small, it is most probable to lose zero or one photons, in which case we would be able to correct the codewords. When α becomes larger, the probability to lose two or three photons, contributing to the set of uncorrectable errors, increases, and at some point, the weights of the corresponding terms will dominate in the density operator, so the fidelity will then drop. However, if α keeps increasing, then the probability to lose four or five photons will also increase and start contributing. This time, because we can correct the errors coming from four and five photon losses, the fidelities will go up again. The situation is similar when α becomes even larger and so the fidelities will oscillate, going up and down. For the case when the repeater stations are included, we also have to include the entanglement swapping operations. According to Section 3.1, $n_e$ (the number of elementary "units") is always even (10, 100, 1000…) except when $L_0 = 1000km$ ($n_e = 1$ in this case), and then according to Equation S21, the fidelities can no longer go below 1/2.

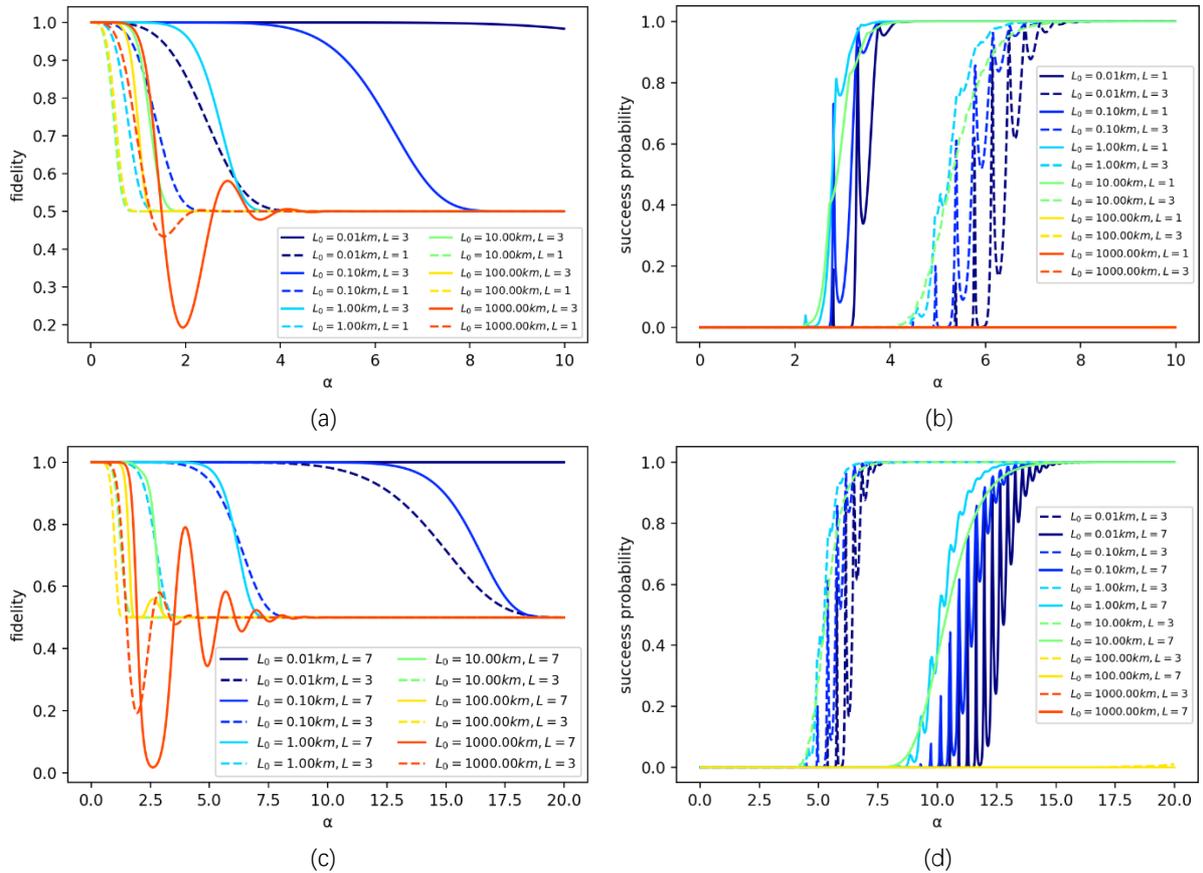

**Figure 4.** The total fidelity $F_{tot}$ ((a), (c)) and the total success probability $P_{tot}$ ((b), (d)) as a function of $\alpha$ for various elementary distances $L_0$: $L_0 = 0.01km$ (dark blue), $L_0 = 0.1km$ (blue), $L_0 = 1km$ (cyan), $L_0 = 10km$ (green), $L_0 = 100km$ (yellow), $L_0 = 1000km$ (orange). We compare them for the 1-loss (dashed) and 3-loss (solid) codes in (a) and (b) and for the 3-loss (dashed) and 7-loss (solid) codes in (c) and (d). The total distance is always chosen to be $1000km$.

To summarize, we need to determine an $\alpha$-regime which allows for near-unit $F_{tot}$ and $P_{tot}$ with $L_0$ being as large as possible. As shown in Figures 4(a) and 4(b), for the 1-loss cat code, this kind of $\alpha$ regime does not exist with $L_0 \geq 0.01km$, which means more than a hundred thousand stations are needed for a total distance of $1000 km$. Obviously, it is then not an optimal choice to use the 1-loss cat code. Fortunately, if we consider higher-loss codes, there is such an $\alpha$ regime with a suitable choice of $L_0$. The behaviors of $F_{tot}$ and $P_{tot}$ based on the 3-loss and 7-loss codes are shown in Figures 4(c) and 4(d). For the 3-loss code, the appropriate $\alpha$ regime does exist, with $L_0 \sim 0.01 km$, and for the 7-loss code, it works with $L_0 \sim 0.1 km$. Thus, one may expect that there is such a suitable $\alpha$ regime for even larger $L_0$ if the loss order goes beyond seven (while $L_0$ must remain sufficiently small; clearly smaller than $15 km$ which corresponds to a channel transmission of at least ½ per segment).

**3.4. Secret Key Rate Analysis**

An important possible application of entanglement distribution is quantum key distribution (**QKD**). Our repeater protocol, in principle, can then be used for long-range QKD. According to Section 3.2 (also see S4), one possibility for the final quantum state (shared among Alice and Bob) is presented in Equation S16. Here we rewrite it as

$$\left(\frac{1}{2} + \frac{1}{2}\prod_{i=0}^{2^m-1}\left(\frac{p_i - p_{i+2^m}}{p_i + p_{i+2^m}}\right)^{t_i}\right)|\Phi^+_{q\pi/2^m}\rangle\langle\Phi^+_{q\pi/2^m}|$$
$$+\left(\frac{1}{2} - \frac{1}{2}\prod_{i=0}^{2^m-1}\left(\frac{p_i - p_{i+2^m}}{p_i + p_{i+2^m}}\right)^{t_i}\right)|\Phi^-_{q\pi/2^m}\rangle\langle\Phi^-_{q\pi/2^m}|, \quad (37)$$

with a corresponding success probability as described in Equation S20. Note here that $|\Phi^\pm_{q\pi/2^m}\rangle$ can also be replaced by $|\Psi^\pm_{q\pi/2^m}\rangle$ depending on all the results of the Bell measurements being done in the entanglement swapping process. If we define the fidelity in this case as $F_{\{t_i\}}$, then Equation 37 can be rewritten as

$$F_{\{t_i\}}|\Phi^+_{q\pi/2^m}\rangle\langle\Phi^+_{q\pi/2^m}| + (1 - F_{\{t_i\}})|\Phi^-_{q\pi/2^m}\rangle\langle\Phi^-_{q\pi/2^m}|. \quad (38)$$

The (asymptotic) secret key rate is the product of the raw rate and the secret key fraction,[30]

$$R_{QKD} \coloneqq R_{raw} r_\infty. \tag{39}$$

The raw rate $R_{raw}$ in our case is

$$R_{raw} = \frac{1}{t_0} P_{tot}, \tag{40}$$

where $t_0 \sim 10^{-6} s$ is the typical repetition time when light-matter interfaces and couplings are involved, which corresponds to an experimental clock rate $\sim$MHz (recall that our scheme is memoryless and hence independent of the usual waiting times for classical signals in memory-based quantum repeater schemes; unlike all-photonic quantum repeaters, however, clock rates $\sim$GHz will be hard to achieve in our case). In our scheme, when employed for QKD, Alice and Bob can measure their atomic states immediately and the complete classical information about the Bell measurement outcomes from the entanglement swapping at all the repeater stations can be transferred to them at the end with no need to wait for any classical signals at an earlier stage. Thus, the raw rates have no fundamental limitation in theory and here we just choose a typical repetition time in the lab in order to analyze the secret key rates. We consider an entanglement-based version of the BB84 scheme. The BB84 secret key fraction is given by[27,28]

$$r_\infty^{BB84} = 1 - h(e_Z) - h(e_X), \tag{41}$$

where $h(p)$ is the binary entropy function and $e_{X/Z}$ are the quantum bit error rates (QBERs). For the explicit state in Equations 37 and 38, the averaged QBERs $e_X$ and $e_Z$ are[28,30]

$$e_X = 1 - F_{\{t_i\}}, \qquad e_Z = 0. \tag{42}$$

After some additional derivations (see S5), we get a lower bound of the averaged secret fraction for the BB84 protocol, which can be written as

$$r_\infty^{BB84} > 1 - h(F_{tot}). \tag{43}$$

Hence the lower bound of the secret key rate is given by

$$R_{QKD} > R_{LB} = \frac{1}{t_0} P_{tot} (1 - h(F_{tot})). \tag{44}$$

The secret key rate incorporates both the fidelity and the success probability. As shown in **Figures 5(a)** and **5(b)**, for the 1-loss code, if the elementary distance is 0.01km, it is possible to get a nonzero secret key rate, but in a rather narrow range of $\alpha$ (total distance: $1000km$). If the elementary distance is equal to 0.1km, the secret key rate is always zero. This is similar to what we discussed in Section 3.3—for $L = 1$, there is no such $\alpha$-regime that both $F_{tot}$ and $P_{tot}$ are large enough. Analogously, for $L = 3$ and $L = 7$, the secret key rate can be nonzero for an elementary distance of 0.01km and 0.1km, but this requires larger $\alpha$. It can be seen from Figures 5(b), 5(d), and 5(f) that the secret key rate per channel use $R_{QKD}t_0$ is much smaller than one with an elementary distance of 0.01km for the 1-loss code. However, it can be very close to unity for the 3-loss code with an elementary distance of 0.01km, and in this case, it can also be nonzero with an elementary distance of 0.1km, but the rate is still not very large in this case. For the 7-loss code, it can be close to unity even with an elementary distance of 0.1km. Therefore, it is reasonable to suppose that for higher-loss codes ($L > 7$), the secret key rate can be nonzero with larger elementary distances and the secret key rate per channel use may even reach unity with a large enough elementary distance (e.g., $L_0 = 0.1km$ or even larger).

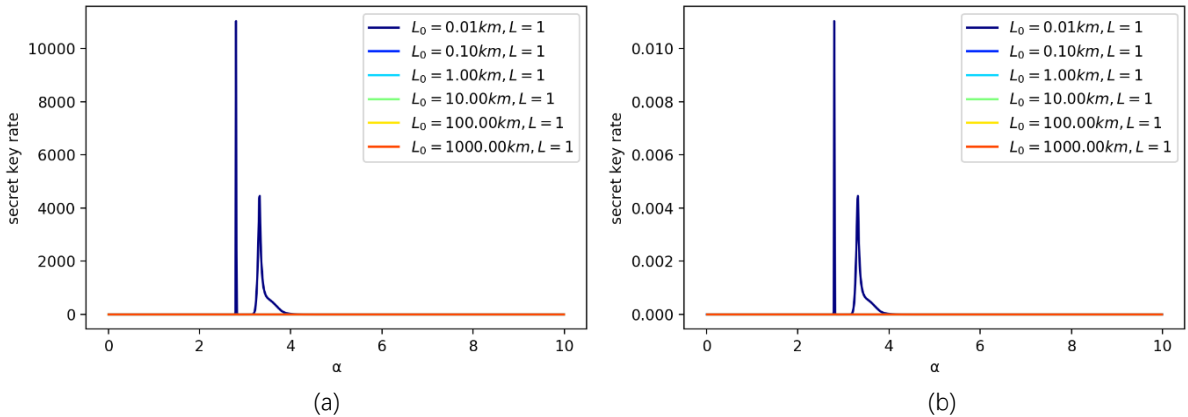

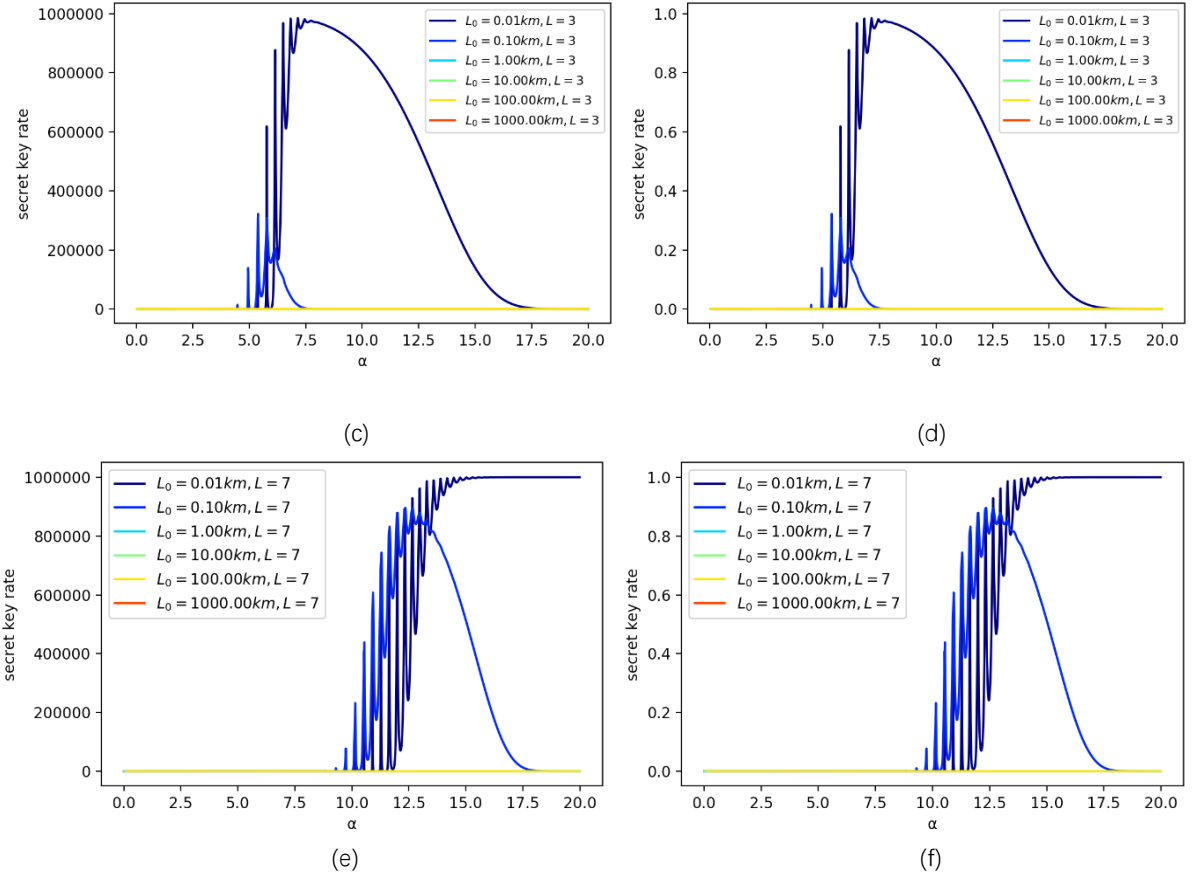

**Figure 5.** The lower bound of the secret key rate per second $R_{QKD}$ ((a), (c), (e)) and, also explicitly shown for convenience, the secret key rate per channel use $R_{QKD} t_0$ ((b), (d), (f)) as a function of $\alpha$ for codes with different loss order: 1-loss code in (a), (b), 3-loss code in (c), (d), 7-loss code in (e), (f), and also for various elementary distances $L_0$: $L_0 = 0.01 km$ (dark blue), $L_0 = 0.1 km$ (blue), $L_0 = 1 km$ (cyan), $L_0 = 10 km$ (green), $L_0 = 100 km$ (yellow), $L_0 = 1000 km$ (orange). The total distance is always chosen to be $1000 km$.

Here, the total distance is always $L_{tot} = 1000 km$ and the corresponding (repeaterless) total transmittance is $\eta_{tot} = \exp(-L_{tot}/L_{att}) \approx 1.82 \times 10^{-20}$, so for the corresponding PLOB bound we have,[31,32] $-\log_2(1 - \eta_{tot}) \approx \eta_{tot}/\ln(2) \approx 2.62 \times 10^{-20}$. Then, as it can be seen in Figure 5, the secret key rates per channel use can significantly exceed the PLOB bound with loss orders $L = 1, 3, 7$ and suitably chosen values for $\alpha$.

Besides channel loss, in a real experiment, there are also errors caused by the local imperfections such as the photon loss related with the cavity interfaces at every repeater station as well as imperfect measurements. What we discussed in Section 2.4 is also applicable for the local photon loss, but the corresponding transmission parameter $\eta$ will depend on the cavities and the way to do the measurements (instead of the channel distance). Since the local errors will accumulate at every station, more stations do not necessarily provide a larger secret key rate in this case. However, note that thanks to the overall avoidance of storing qubits, there are no extra memory errors in our scheme such as the usual memory dephasing in memory-based quantum repeaters. Also note that our model to include the local imperfections exclusively via local losses means that also the finally shared state in Equations 37 and 38 maintains its specific form where only one type of Pauli errors occurs resulting in a nonzero QBER only in one of the two BB84 variables (see Equation 42). In a more general model that takes into account other local operational errors such as qubit depolarizing errors, both QBERs become nonzero and in this case the secret key fraction may sharply drop to zero for insufficient depolarization error parameters even when all the other parameters take on very good values (see Equation 41). However, in principle, also in this more general setting, general quantum error correction codes can provide a certain level of protection against such local errors. Experimentally, both local loss and local (dephasing or depolarizing) noise may occur with typically one dominating over the other.

In order to evaluate the effects caused by the local losses, we set $\eta_{local} = 0.99, 0.999$. In **Figures 6(c)** and **6(e)**, one can see that the secret key rates are indeed reduced, but they would still overcome the PLOB bound, even when the local errors are included with a loss order of $L = 3, 7$. For $\eta_{local} = 0.999$, the rates do not change a lot and we can still realize near-unit rates per channel use for certain values of $\alpha$ and $L_0$. On the other hand, for $\eta_{local} = 0.99$, the final rates decrease greatly. However, we can see in Figures 6(d) and 6(f), even though the rates are much smaller than unity, they can still beat the PLOB bound. Also, the peaks are very narrow for $\eta_{local} = 0.99$, so to reach the optimal secret key rates, one needs to have a pretty accurate control of the value of $\alpha$. As for the 1-loss code, we can see in Figures 6(a) and 6(b) that the rates already decrease significantly even for $\eta_{local} = 0.999$ and there is only a really narrow peak to overcome the PLOB bound. For $\eta_{local} = 0.99$, the rates cannot beat the PLOB bound even with very short $L_0$. Moreover, one can see that a shorter elementary distance $L_0$ does not necessarily bring us a better performance of the secret key rate according to Figures 6(a) and 6(b). In Figure 6(a), it can be seen that the best choice for $L_0$ is not the shortest one, $L_0 = 0.01km$, but instead a larger one, $L_0 = 0.1km$, and in Figure 6(b), the best choice is even larger than that—$L_0 = 1km$.

The secret key rate inevitably drops if the local losses are included in addition to the channel losses and a consequence is that the local losses must not be too large ( $\eta_{local} \geq 0.999$ ) in order to obtain an $\alpha$-regime where the secret key rate per channel use is close to unity. However, as already mentioned in Section 3.3, even though the numerical analysis for higher-loss codes beyond $L = 7$ has not been done yet, we expect that if the loss order $L$ could be larger ($L > 7$), there should be an $\alpha$-regime with reasonable values of $L_0$ (note again that the shortest $L_0$ may not be the best choice) and also the maximally allowed local losses could be a bit larger (e.g., $\eta_{local} \leq 0.99$). Nonetheless, keeping the local imperfections at very low values ($10^{-3}$- $10^{-2}$ or possibly $10^{-1}$ corresponding to a regime less demanding than for the typical threshold values in fault-tolerant quantum computing) is a requirement of our scheme which is not easy to meet in a practical realization.

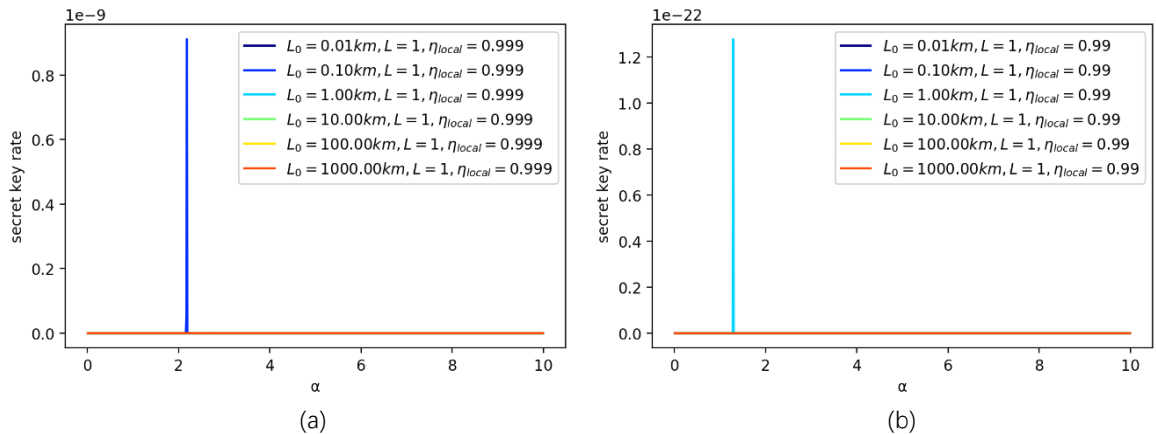

(a)        (b)

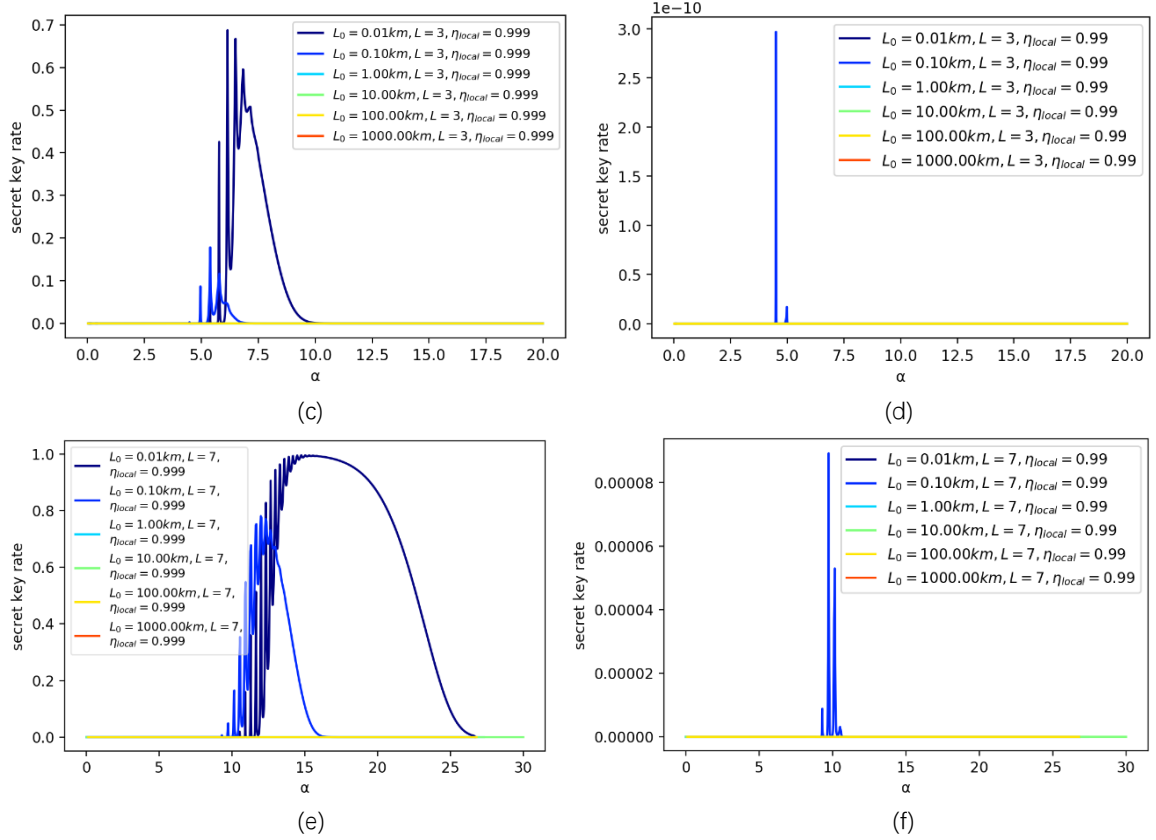

**Figure 6.** The lower bound of the secret key rate per channel use $R_{QKD}t_0$ including local loss: $\eta_{local} = 0.999$ ((a), (c), (e)) and $\eta_{local} = 0.99$ ((b), (d), (f)) as a function of $\alpha$ for codes with different loss orders: 1-loss code in (a), (b), 3-loss code in (c), (d), 7-loss code in (e), (f), and also for various elementary distances $L_0$: $L_0 = 0.01km$ (dark blue), $L_0 = 0.1km$ (blue), $L_0 = 1km$ (cyan), $L_0 = 10km$ (green), $L_0 = 100km$ (yellow), $L_0 = 1000km$ (orange). The total distance is always chosen to be $1000km$.

## 4. Conclusion

We proposed a scheme for long-distance entanglement distribution based on so-called 'rotation-symmetric bosonic codes' (RSBCs) and cavity-QED. The cavity-QED part employs a three-level atom confined in a high-finesse cavity. An exploitation of the atomic system for quantum storage is unnecessary, because our protocol, though based on spin-spin entanglement distribution, does not involve any extra classical signaling times for synchronization of the repeater segments, utilizing quantum error correction of photon loss errors. We showed that the cavity-QED-based approach can be used to generate the RSBC for the propagating light mode, do the syndrome measurement, and create the entanglement. Thanks to the properties of such codes, the dominating errors induced from photon losses along the optical fiber channel can be immediately corrected at the intermediate repeater stations.

A specific instance of a RSBC—the cat code has been analyzed in detail. To get a better state fidelity, generally higher-loss codes are needed. However, in this case, the overlap of the codewords increases, which results in a worse success probability of the optical, unambiguous state discrimination (USD), as employed in our scheme to disentangle the light modes from the entangled spin states. Fortunately, we found that for higher-loss codes, fewer elementary repeater stations are required in order to get an $\alpha$ regime ($\alpha$ is the light mode amplitude) where both the final state fidelity and the overall success probability become near-unity. For QKD applications, the secret key rate analysis leads to a similar conclusion, namely that for higher-loss codes, fewer elementary repeater stations are needed to obtain an $\alpha$ regime where the secret key rate per channel use is close to unity. Final secret key rates of the order of MHz are then possible, i.e., in principle, final long-distance quantum communication rates that match the experimentally given clock rates for cavity-QED atom-light systems. However, once the local losses at all the repeater stations are included into the analysis, as required to assess a more realistic scenario, the key rate can drop dramatically for a local transmission parameter chosen too small (i.e., smaller than 0.99). Hence, really small local losses (with a local transmission greater than 0.999) must be assumed at every elementary station in our scheme to keep the key rate close to unity. Nonetheless, it is also predicted that even higher-loss codes than those considered here (i.e., $L > 7$) may lead to a further improvement.

Cat qubits could also be embedded into a higher multi-qubit code,[11, 18, 19] e.g., the so-called quantum parity code, combining many physical qubits into a logical qubit. Unlike the usual single-photon dual-rail (two-mode) qubits for which some physical qubits are considered to be erased through the loss channel (i.e. two independently acting, identical oscillator amplitude damping channels), the single-mode qubits in bosonic, continuous-variable codes will not be naturally erased in a fiber channel. Instead, in this case, the single-mode amplitude damping generally leads to a distortion of the quantum states. Nonetheless, we may choose to discard some qubits artificially, e.g., we can do the USD on one of the physical qubits and if it turns out to be unsuccessful, we then discard this qubit and attempt the USD for the remaining physical qubits. So, with this method, in principle, we can have a higher chance to accomplish the USD and the overall success probability of such a repeater scheme will be improved. A crucial benefit of our current approach though is that it is hardware-efficient and thus no additional multi-mode couplings to generate the higher-order multi-qubit codes are needed.

There are still various other kinds of RSBCs, e.g., those based on squeezed cat states, binomial states, or Pegg-Barnett states.[8] Our scheme for the syndrome measurement is expected to work also for the generalized RSBCs. However, the performance in terms of fidelity is no longer so clear for the other RSBCs, but the overlap of the codewords potentially becomes smaller (so the success probability is expected to be improved) than for the cat codes, especially when the squeezed cat codes and the binomial codes are considered. Thus, as a combination of the final state fidelity and the overall success probability, the secret key rate of long-range QKD based on our repeater scheme should also be improved with this generalization. To conclude, the other RSBCs potentially have a better performance than the cat codes and we leave it for future work to analyze the performance of other instances of RSBCs. Generally, we hope that our scheme will inspire further experimental research towards realizing elements of the proposed method with the ultimate goal of long-distance quantum communication. For a practical, real-world application, however, the local losses and errors that occur at every repeater station in our scheme are a serious obstacle and so further loss/error-suppressing elements would have to be incorporated into our scheme, for instance, via additional quantum error correction codes on the atomic spin qubits.

**Acknowledgement**

We thank the BMBF in Germany for support via Q.Link.X/QR.X and the BMBF/EU for support via QuantERA/ShoQC.

# Supporting Information

**Memoryless quantum repeaters based on cavity-QED and coherent states**

*Pei-Zhe Li and Peter van Loock\**

**S1. Proof of the Syndrome Measurement Process**

Here we use mathematical induction to prove that the syndrome measurement process works for any $\phi \geq \pi/2^{m-1}$. First, the process works for $\phi = \pi$ and $\phi = \pi/2$ according to Section 2.2. Then assuming that it works for $\phi = \pi/2^k, 1 \leq k \leq m-1 \ \& \ k \in \mathbb{N}$, i.e., we know the remainder of $q$ divided by $2^{k+1}$. We define

$$q = j \cdot 2^{k+1} + r_{2^{k+1}}(q), \tag{S1}$$

where $j$ is an integer and $r_{2^{k+1}}(q)$ is the remainder. Then we have

$$q = \frac{j}{2} \cdot 2^{k+2} + r_{2^{k+1}}(q) = \frac{j-1}{2} \cdot 2^{k+2} + \left(r_{2^{k+1}}(q) + 2^{k+1}\right). \tag{S2}$$

Thus, there are only two possibilities of the remainder of $q$ divided by $2^{k+2}$,

$$r_{2^{k+2}}(q) = r_{2^{k+1}}(q), \quad \text{if } j \text{ is even,}$$

$$r_{2^{k+2}}(q) = r_{2^{k+1}}(q) + 2^{k+1}, \quad \text{if } j \text{ is odd.} \tag{S3}$$

Now considering the light-atom interaction, this time we prepare the cavity with $\phi = \pi/2^{k+1}$, then according to Section 2.2, the state after the reflection is

$$\frac{1}{2}\left(|\uparrow\rangle_s + e^{-\frac{iq\pi}{2^{k+1}}}|\downarrow\rangle_s\right)\left(|\uparrow\rangle_A \hat{a}^q |0_{2^m,\Theta}\rangle + |\downarrow\rangle_A \hat{a}^q |1_{2^m,\Theta}\rangle\right), \tag{S4}$$

where the phase is constrained to two possible values according to Equation S3, i.e.,

$$e^{-\frac{iq\pi}{2^{k+1}}} = e^{-\frac{ir_{2^{k+1}}(q)\pi}{2^{k+1}}}, \quad \text{if } j \text{ is even,}$$

$$e^{-\frac{iq\pi}{2^{k+1}}} = e^{-\left(\frac{ir_{2^{k+1}}(q)}{2^{k+1}}+1\right)\pi} = -e^{-\frac{ir_{2^{k+1}}(q)\pi}{2^{k+1}}}, \quad \text{if } j \text{ is odd.} \tag{S5}$$

It is obvious that the states $|\uparrow\rangle_s + \exp(i\psi)|\downarrow\rangle_s$ and $|\uparrow\rangle_s - \exp(i\psi)|\downarrow\rangle_s$ are orthogonal for arbitrary $\psi$. Therefore, the two possibilities of the atomic state can be distinguished

deterministically meaning that the remainder of $q$ divided by $2^{k+2}$ is known. In conclusion, we can finally get the remainder of $q$ divided by $2^m$ for arbitrary integer $m$ using the method mentioned in Section 2.2.

## S2. Alternative Way for Experiments

Thanks to the symmetric properties of the RSBCs, all the methods mentioned can be replaced with the rotation angle $\pi/2 \leq \phi \leq \pi$, which is more practical in experiments. This can be realized simply by replacing the $\phi$ mentioned above by $\pi - \phi$ unless $\phi = \pi$. According to Equation 19, the bit flip happens if we perform a rotation operation on $|0_{2^m,\Theta}\rangle$ and $|1_{2^m,\Theta}\rangle$ with $\phi = \pi/2^m$. Now we look at the results when $\phi = \pi - \pi/2^m$,

$$\hat{R}\left(\pi - \frac{\pi}{2^m}\right)|0_{2^m,\Theta}\rangle = \hat{R}(\pi)\hat{R}\left(-\frac{\pi}{2^m}\right)|0_{2^m,\Theta}\rangle$$

$$= \hat{R}(\pi)\hat{R}\left(-\frac{\pi}{2^m}\right)\hat{R}\left(\frac{\pi}{2^{m-1}}\right)|0_{2^m,\Theta}\rangle$$

$$= \hat{R}(\pi)\hat{R}\left(\frac{\pi}{2^m}\right)|0_{2^m,\Theta}\rangle$$

$$= \hat{R}(\pi)|1_{2^m,\Theta}\rangle = |1_{2^m,\Theta}\rangle. \tag{S6}$$

It is analogous if we exchange $|0_{2^m,\Theta}\rangle$ and $|1_{2^m,\Theta}\rangle$ and here we make use of Equation 19 and Equation 15. Thus, the results of rotation operator $\hat{R}(\pi - \pi/2^m)$ acting on the two corresponding codewords $|0_{2^m,\Theta}\rangle$ and $|1_{2^m,\Theta}\rangle$ are the same as the rotation operator $\hat{R}(\pi/2^m)$ meaning that all the cavities with the rotation angle $\phi$ prepared for generating codes and creating entanglement can be replaced by those with $\pi - \phi$.

For the syndrome measurement, first it is easy to see from Equation 15 that

$$\hat{R}\left(\pi - \frac{\pi}{2^{m-1}}\right)|0_{2^m,\Theta}\rangle = |0_{2^m,\Theta}\rangle,$$

$$\hat{R}\left(\pi - \frac{\pi}{2^{m-1}}\right)|1_{2^m,\Theta}\rangle = |1_{2^m,\Theta}\rangle. \tag{S7}$$

Then Equation 13 for the joint atom-light state becomes

$$\frac{1}{2}\left(|\uparrow\rangle_s + e^{iq\left(\frac{\pi}{2^{m-1}}-\pi\right)}|\downarrow\rangle_s\right)\left(|\uparrow\rangle_A \hat{a}^q |0_{2^m,\Theta}\rangle + |\downarrow\rangle_A \hat{a}^q |1_{2^m,\Theta}\rangle\right). \tag{S8}$$

Here the phase is

$$e^{iq\left(\frac{\pi}{2^{m-1}}-\pi\right)} = \begin{cases} e^{\frac{iq\pi}{2^{m-1}}}, & \text{if } q \text{ is even,} \\ -e^{\frac{iq\pi}{2^{m-1}}}, & \text{if } q \text{ is odd.} \end{cases} \tag{S9}$$

Since there is no such replacement if $\phi = \pi$, we already know $q$ is even or odd immediately after the first step. Therefore, all the steps remain the same as described in Section 2.2 and S1. Only one thing needs to be kept in mind, namely that one needs to be aware that the phase now is $\exp(iq\pi/2^{m-1})$ or $-\exp(iq\pi/2^{m-1})$ depending on the parity of $q$ instead of $\exp(-iq\pi/2^{m-1})$ when analyzing the results from experiments.

**S3. Analysis for the Orthogonal Codewords**

The codewords as discussed in Section 2 can also be replaced by orthogonal ones. The orthogonal codewords can be written as,[S1]

$$|0_{M,\Theta}\rangle_\perp = \frac{1}{\sqrt{\mathcal{N}_{M,0}}} \sum_{k=0}^{2M-1} e^{\frac{k\pi i}{M}\hat{n}} |\Theta\rangle,$$

$$|1_{M,\Theta}\rangle_\perp = \frac{1}{\sqrt{\mathcal{N}_{M,1}}} \sum_{k=0}^{2M-1} (-1)^m e^{\frac{k\pi i}{M}\hat{n}} |\Theta\rangle. \tag{S10}$$

Then it is straightforward to get the following relation between the nonorthogonal codewords and the orthogonal ones,

$$|0_{M,\Theta}\rangle_\perp = \frac{1}{\sqrt{\mathcal{N}'_{M,0}}}\left(|0_{M,\Theta}\rangle + |1_{M,\Theta}\rangle\right)$$

$$|1_{M,\Theta}\rangle_\perp = \frac{1}{\sqrt{\mathcal{N}'_{M,1}}}\left(|0_{M,\Theta}\rangle - |1_{M,\Theta}\rangle\right). \tag{S11}$$

where the normalization factors are given by $\mathcal{N}'_{M,0} = N_M/\mathcal{N}_{M,0}$, $\mathcal{N}'_{M,1} = N_M/\mathcal{N}_{M,1}$.

The orthogonal codewords can be generated using the method presented in Section 2.1. The Equation 9 can be rewritten in the basis $\{|\pm\rangle\}$ just like Equation 7, so $|0_{2^m,\Theta}\rangle_\perp$ or $|1_{2^m,\Theta}\rangle_\perp$ can be prepared by doing the measurement in the basis $\{|\pm\rangle\}$. However, according to Equation S11, the normalization factors before $|0_{2^m,\Theta}\rangle_\perp$ and $|1_{2^m,\Theta}\rangle_\perp$ in the rewritten expression are different. In fact, it is not possible to get a joint atom-light state with balanced norms with $|0_{2^m,\Theta}\rangle_\perp$ and $|1_{2^m,\Theta}\rangle_\perp$ like $|0_{2^m,\Theta}\rangle$ and $|1_{2^m,\Theta}\rangle$ in Equation 9 using this particular method. The syndrome measurement can also be done on these orthogonal codewords, but for the entanglement creation process, the final state should be (code space)

$$\frac{1}{2}\left(|+\rangle_B|+\rangle_A \sqrt{\mathcal{N}'_{M,0}}|0_{2^m,\Theta}\rangle_\perp + |-\rangle_B|-\rangle_A \sqrt{\mathcal{N}'_{M,1}}|1_{2^m,\Theta}\rangle_\perp\right). \qquad (S12)$$

This is a GHZ-type state,[S2] so in order to get the states in Alice and Bob's hands entangled, the measurement on the light mode should be done in the basis

$\left(\sqrt{\mathcal{N}'_{M,0}}|0_{2^m,\Theta}\rangle_\perp + \sqrt{\mathcal{N}'_{M,1}}|1_{2^m,\Theta}\rangle_\perp\right)/2 = |0_{M,\Theta}\rangle$ and

$\left(\sqrt{\mathcal{N}'_{M,0}}|0_{2^m,\Theta}\rangle_\perp - \sqrt{\mathcal{N}'_{M,1}}|1_{2^m,\Theta}\rangle_\perp\right)/2 = |1_{M,\Theta}\rangle$, which then just goes back to the discrimination of the nonorthogonal codewords. Therefore, even if we use the orthogonal codewords in transmission, the success probability to create an entangled pair is still not unity. In fact, Equation S12 can be rewritten as a special case of Equation 20 with $q = 0$, so it is not surprising that the success probability is not improved. The main problem is the different norms of $|0_{2^m,\Theta}\rangle_\perp$ and $|1_{2^m,\Theta}\rangle_\perp$ in Equation S12 (as it can also be seen from Equation 7). In fact, it is not possible to generate the states with the same norms with this particular method. However, even if it can be realized in other ways, the orthogonal codewords will lead to a deformation of the logical qubits, as soon as loss is included.[S3] This deformation will certainly reduce the fidelity and might also make the entanglement creation process more subtle. In conclusion, we only consider the nonorthogonal codewords throughout.

## S4. Analysis for Entanglement Swapping

The entanglement swapping is the essential part in our protocol to distribute entanglement over a long distance. In our scheme, an entangled pair is generated at every elementary unit with some errors. For a code with loss order $L = 2^m - 1$, according to Section 3.2, the final state at every elementary unit is shown in Equation 32 for a given $q$ with corresponding probability $p_q + p_{q+2^m}$. After the discrimination between $|0^q_{2^m,\Theta}\rangle$ and $|1^q_{2^m,\Theta}\rangle$, the atoms in $QR_1$ and $QR_2$ are entangled and the corresponding density operator for a given $q$ is

$$\hat{\rho}^p_{1,2} = \begin{cases} \dfrac{p_q}{p_q + p_{q+2^m}} |\Phi^+_{q\pi/2^m}\rangle\langle\Phi^+_{q\pi/2^m}| + \dfrac{p_{q+2^m}}{p_q + p_{q+2^m}} |\Phi^-_{q\pi/2^m}\rangle\langle\Phi^-_{q\pi/2^m}|, \\ \qquad LM \text{ in } |0^q_{2^m,\Theta}\rangle, \\ \dfrac{p_q}{p_q + p_{q+2^m}} |\Psi^+_{q\pi/2^m}\rangle\langle\Psi^+_{q\pi/2^m}| + \dfrac{p_{q+2^m}}{p_q + p_{q+2^m}} |\Psi^-_{q\pi/2^m}\rangle\langle\Psi^-_{q\pi/2^m}|, \\ \qquad LM \text{ in } |1^q_{2^m,\Theta}\rangle, \end{cases} \qquad (S13)$$

where $LM$ represents the light mode. Since we know which form the density operator takes on according to the results from the USD, the two forms are, in fact, equivalent for the contribution of the fidelity.

In order to do the entanglement swapping, a joint Bell measurement is required and the final entangled state after the swapping depends on the result of the Bell measurement. According to Ref. [S4], for two identical input states, we assume the input-state density operator for both of them is

$$\hat{\sigma}_1 = A|\Phi^+_{q\pi/2^m}\rangle\langle\Phi^+_{q\pi/2^m}| + B|\Phi^-_{q\pi/2^m}\rangle\langle\Phi^-_{q\pi/2^m}|. \qquad (S14)$$

Then after the swapping, the state becomes

$$\begin{pmatrix} \frac{1}{2} & 0 & 0 & \frac{1}{2}(A-B)^2 \\ 0 & 0 & 0 & 0 \\ 0 & 0 & 0 & 0 \\ \frac{1}{2}(A-B)^2 & 0 & 0 & \frac{1}{2} \end{pmatrix} \qquad (S15)$$

when $|\Phi^+_{q\pi/2^m}\rangle$ is the result of the Bell measurement. For two identical states whose density operator is

$$\hat{\sigma}_2 = C|\Psi^+_{q\pi/2^m}\rangle\langle\Psi^+_{q\pi/2^m}| + D|\Psi^-_{q\pi/2^m}\rangle\langle\Psi^-_{q\pi/2^m}|, \qquad (S16)$$

the subsequent state after swapping is

$$\begin{pmatrix} 0 & 0 & 0 & 0 \\ 0 & \frac{1}{2} & \frac{1}{2}(C-D)^2 & 0 \\ 0 & \frac{1}{2}(C-D)^2 & \frac{1}{2} & 0 \\ 0 & 0 & 0 & 0 \end{pmatrix} \qquad (S17)$$

when $|\Psi^+_{q\pi/2^m}\rangle$ is the result of the Bell measurement. Next we consider that the two input states are different, i.e., the density operator of one of them is $\hat{\sigma}_1$ and the other one is $\hat{\sigma}_2$. Then after the swapping, the state becomes

$$\begin{pmatrix} \frac{1}{2} & 0 & 0 & \frac{1}{2}(A-B)(C-D) \\ 0 & 0 & 0 & 0 \\ 0 & 0 & 0 & 0 \\ \frac{1}{2}(A-B)(C-D) & 0 & 0 & \frac{1}{2} \end{pmatrix} \qquad (S18)$$

when $|\Phi^+_{q\pi/2^m}\rangle$ is the result of the Bell measurement. If the results of the Bell measurement are different from what was mentioned above, the subsequent states are similar and the fidelities remain the same. From Equations S15, S17 and S18, one can find that the fidelity after the swapping only depends on the corresponding coefficients ($A$, $B$, $C$ and $D$) and it remains the same independent of whether the input state is a mixture of $|\Phi^+_{q\pi/2^m}\rangle$ and $|\Phi^-_{q\pi/2^m}\rangle$ or a mixture of $|\Psi^+_{q\pi/2^m}\rangle$ and $|\Psi^-_{q\pi/2^m}\rangle$. An instance for the density matrix of the final state after all the entanglement swapping processes is

$$\begin{pmatrix} \frac{1}{2} & 0 & 0 & \frac{1}{2}\prod_{i=0}^{2^m-1}\left(\frac{p_i - p_{i+2^m}}{p_i + p_{i+2^m}}\right)^{t_i} \\ 0 & 0 & 0 & 0 \\ 0 & 0 & 0 & 0 \\ \frac{1}{2}\prod_{i=0}^{2^m-1}\left(\frac{p_i - p_{i+2^m}}{p_i + p_{i+2^m}}\right)^{t_i} & 0 & 0 & \frac{1}{2} \end{pmatrix} \qquad (S19)$$

with probability

$$p_{\{t_i\}} = \frac{n!}{t_0! \, t_1! \cdots t_{2^m-1}!} \prod_{i=0}^{2^m-1} (p_i + p_{i+2^m})^{t_i}. \tag{S20}$$

Equation S20 is one of the terms in the expansion of $(p_0 + p_1 + \cdots + p_{2^{m+1}-1})^{n_e}$. The sum of all $t_i$s is $n_e$ and $\{t_i\}$ represents one possible combination of the integers $t_0$ to $t_{2^m-1}$. Thus, the averaged total fidelity is

$$\begin{aligned}
F_{tot} &= \frac{1}{2} \sum \frac{n!}{t_0! \, t_1! \cdots t_{2^m-1}!} \prod_{i=0}^{2^m-1} (p_i + p_{i+2^m})^{t_i} \\
&\quad + \frac{1}{2} \sum \frac{n!}{t_0! \, t_1! \cdots t_{2^m-1}!} \prod_{i=0}^{2^m-1} (p_i + p_{i+2^m})^{t_i} \left( \frac{p_i - p_{i+2^m}}{p_i + p_{i+2^m}} \right)^{t_i} \\
&= \frac{1}{2} + \frac{1}{2} (p_0 - p_{2^m} + p_1 - p_{2^m+1} + \cdots + p_{2^m-1} - p_{2^{m+1}-1})^{n_e} \\
&= \frac{1}{2} + \frac{1}{2} (2F_0 - 1)^{n_e},
\end{aligned} \tag{S21}$$

where the sum is taken over all combinations of $t_0$ to $t_{2^m-1}$.

## S5. Lower Bound of the Secret Key Fraction

The binary entropy is given by

$$h(p) = -p \log_2 p - (1-p) \log_2(1-p), \tag{S22}$$

where $p \in (0,1)$ and $h(p)$ is a concave function in its domain. Thus, for any given $\alpha_i, x_i \in (0,1)$ satisfying $\sum \alpha_i = 1$, we have (Jensen's inequality)

$$\sum_i \alpha_i h(x_i) < h\left( \sum_i \alpha_i x_i \right). \tag{S23}$$

The final state presented in Equation S19 can also be written as

$$\begin{aligned}
&\left( \frac{1}{2} + \frac{1}{2} \prod_{i=0}^{2^m-1} \left( \frac{p_i - p_{i+2^m}}{p_i + p_{i+2^m}} \right)^{t_i} \right) |\Phi^+_{q\pi/2^m}\rangle\langle\Phi^+_{q\pi/2^m}| \\
&+ \left( \frac{1}{2} - \frac{1}{2} \prod_{i=0}^{2^m-1} \left( \frac{p_i - p_{i+2^m}}{p_i + p_{i+2^m}} \right)^{t_i} \right) |\Phi^-_{q\pi/2^m}\rangle\langle\Phi^-_{q\pi/2^m}|
\end{aligned} \tag{S24}$$

with probability described as in Equation S20. Note here $|\Phi^{\pm}_{q\pi/2^m}\rangle$ can also be replaced by $|\Psi^{\pm}_{q\pi/2^m}\rangle$ depending on all the results of the Bell measurements being done in the entanglement swapping process. If we define the fidelity in this case as $F_{\{t_i\}}$, then Equation S16 can be rewritten as

$$F_{\{t_i\}}|\Phi^{+}_{q\pi/2^m}\rangle\langle\Phi^{+}_{q\pi/2^m}| + (1 - F_{\{t_i\}})|\Phi^{-}_{q\pi/2^m}\rangle\langle\Phi^{-}_{q\pi/2^m}|. \qquad (S25)$$

Therefore, according to Section 3.4, the averaged secret fraction for the BB84-protocol is

$$r_{\infty}^{BB84} = \sum p_{\{t_i\}}\left(1 - h(1 - F_{\{t_i\}})\right). \qquad (S26)$$

Here $\{t_i\}$ represents one possible combination of the integers $t_0$ to $t_{2^m-1}$ and the sum is taken over all combinations. Then a lower bound of $r_{\infty}^{BB84}$ can be derived from Equation S23,

$$r_{\infty}^{BB84} = 1 - \sum p_{\{t_i\}}h(F_{\{t_i\}}) > 1 - h\left(\sum p_{\{t_i\}}F_{\{t_i\}}\right) = 1 - h(F_{tot}). \qquad (S27)$$

**S6. A Linear-Optics Solution for the USD of Cat Codes**

The state discrimination plays an important role in our protocol. To get the atomic states entangled, it is necessary to do the state discrimination for the light mode. Since the states to be discriminated in our scheme are not orthogonal, it is not possible, even in theory, to distinguish them deterministically. We used the optimal success probability of the USD throughout, but it is not so clear how we can actually reach the optimal probability. There is a protocol to distinguish the coherent states with different phases only using some beam splitters,[S5] but there is no such general scheme to distinguish cat states. Here we shall propose a possible realization of the USD. This realization cannot reach the optimal probability, but it is not too far from it and only some beam splitters are required for this method.

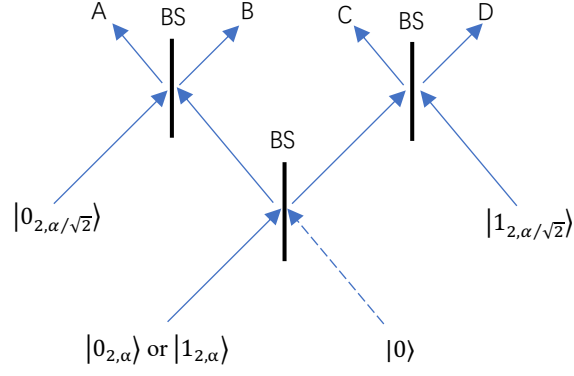

**Figure S1.** A linear-optics solution of the USD for the 1-loss cat code.

For the 1-loss code, the states we need to distinguish are $|0_{2,\alpha}\rangle = |\alpha\rangle + |-\alpha\rangle$ and $|1_{2,\alpha}\rangle = |i\alpha\rangle + |-i\alpha\rangle$.[S6] The first step is to inject the light mode to one port of a beam splitter and inject nothing to the other input port, i.e., the state in that port is the vacuum state $|0\rangle$. Then after the beam splitter interaction, the output state is (first we assume the input state is $|0_{2,\alpha}\rangle$; unnormalized)

$$|\alpha/\sqrt{2}, \alpha/\sqrt{2}\rangle + |-\alpha/\sqrt{2}, -\alpha/\sqrt{2}\rangle. \quad (S28)$$

Then we connect the two output ports of this beam splitter with one of the input ports of two other beam splitters. We prepare the states injected to the other input ports of these two beam splitters as $|0_{2,\alpha/\sqrt{2}}\rangle$ and $|1_{2,\alpha/\sqrt{2}}\rangle$, respectively.[S7] Then after all the interactions, the final state becomes (unnormalized)

$$(|0, \alpha\rangle_{AB} + |\alpha, 0\rangle_{AB}) \otimes$$
$$(|(1-i)\alpha/2, (1+i)\alpha/2\rangle_{CD} + |(1+i)\alpha/2, (1-i)\alpha/2\rangle_{CD}) +$$
$$(|0, -\alpha\rangle_{AB} + |-\alpha, 0\rangle_{AB}) \otimes$$
$$(|(-1-i)\alpha/2, (-1+i)\alpha/2\rangle_{CD} + |(-1+i)\alpha/2, (-1-i)\alpha/2\rangle_{CD}), \quad (S29)$$

where the subscripts, A, B, C, D, correspond to the output ports of the beam splitters as shown in **Figure S1**. Finally, we employ the photon detectors at every output port to measure if photons are present there. From Equation S29, we can see that if the input state is $|0_{2,\alpha}\rangle$, there must be no photon at least at one of the output ports A or B and it is very likely that we can detect photons at both output ports C and D. Similarly, we can find that if the input state is $|1_{2,\alpha}\rangle$, there must be no photon at least at one of the output ports C or D and it is very likely that we can detect photons at both output ports A and B. Thus, we can unambiguously confirm the input state is $|0_{2,\alpha}\rangle$, if we detect photons at both output ports C and D and we can confirm the input state is $|1_{2,\alpha}\rangle$, if we detect photons at both output ports A and B. But if we detect no photons at at least one of the output ports A and B and at the same time, we detect no photons at at least one of the output ports C and D, then it is a failure event and so we cannot determine which is the input state. The corresponding success probability is

$$1 - \frac{2\cos\left(\frac{\alpha^2}{2}\right) + \left(1 + e^{-\alpha^2}\right)\left(\cos(\alpha^2) + e^{\alpha^2} + e^{-\frac{3}{2}\alpha^2}\right)}{2e^{-\alpha^2}\cosh\left(\frac{\alpha^2}{2}\right)\cosh(\alpha^2)}. \quad (S30)$$

There is an example for the comparison between the optimal success probability and the success probability for this particular realization in **Figure S2**. One can see that the success probability presented in Equation S30 is smaller than the optimal one, but the two possibilities are not too far away from each other. This realization can also be extended to higher-loss codes by adding more beam splitters to split the input light mode into more parts.

The secret key rate will certainly decrease with this success probability. For the 1-loss code, as we can see in **Figure S3**, even though the secret key rate decreases greatly, it can still overcome the PLOB bound with $L_0 = 0.01 km$ (here no local errors are considered). We expect that for higher-loss codes, the rates should also be decreased but could still beat the PLOB bound with certain $L_0$ and $\alpha$ and the rates would not decrease too much with a small enough local error (e.g., $\eta_{local} = 0.999$).

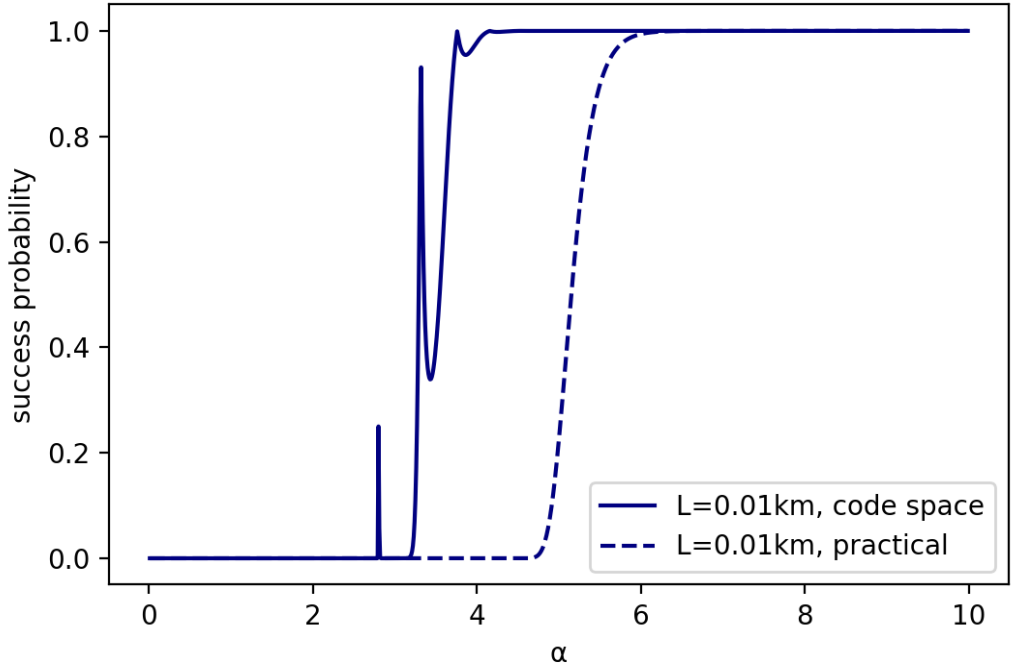

**Figure S2.** The success probability of the USD for the 1-loss cat code. The solid blue line shows the theoretically optimal probability and the dashed blue line is the actual probability using the more "practical" linear-optics solution.

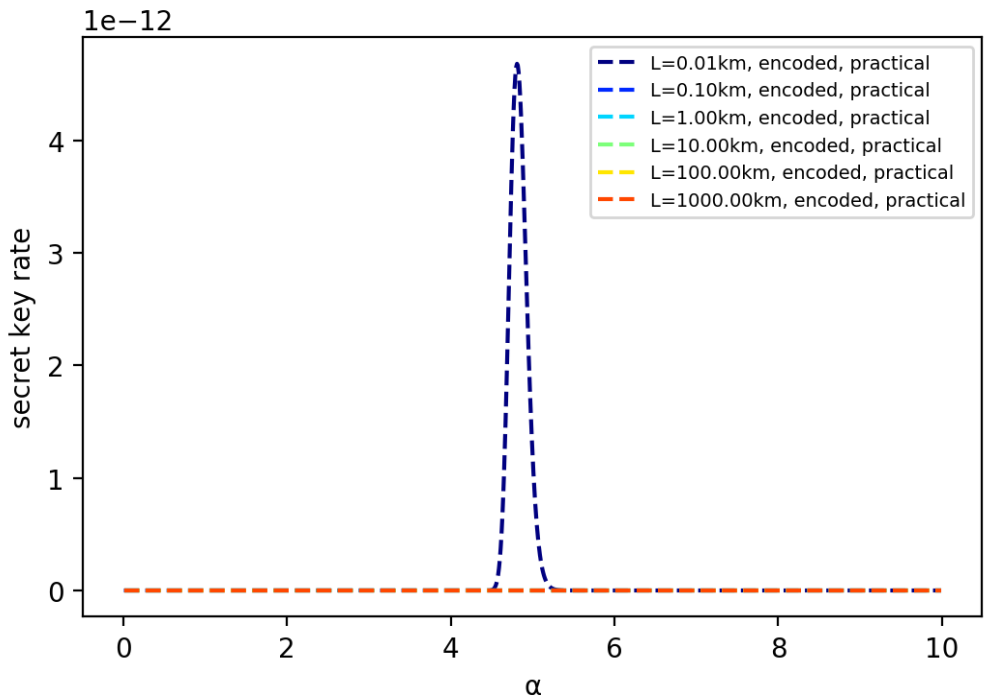

**Figure S3.** The secret key rate per channel use with the linear-optics solution for the 1-loss cat code and for various elementary distances $L_0$: $L_0 = 0.01km$ (dark blue), $L_0 = 0.1km$ (blue), $L_0 = 1km$ (cyan), $L_0 = 10km$ (green), $L_0 = 100km$ (yellow), $L_0 = 1000km$ (orange).

## S7. Entanglement creation without memories

The entangled atomic states can be created without using memories at the elementary stations (but in case they want to keep them, Alice and Bob may use memories to store the entangled pairs). To illustrate this, let us consider the simplest case where there is only one elementary repeater station (this should be the $ES_1$ in Figure 3) between Alice and Bob. Then when the state generation is finished at the two parts of the station, according to Section 2.1, the joint state in the station becomes

$$\frac{1}{\sqrt{2}}\left(|\uparrow\rangle|0_{2^m,\Theta}\rangle + |\downarrow\rangle|1_{2^m,\Theta}\rangle\right)^{\otimes 2}. \tag{S31}$$

It can be rewritten as

$$\frac{1}{\sqrt{2}}\left(|\uparrow\rangle|0_{2^m,\Theta}\rangle + |\downarrow\rangle|1_{2^m,\Theta}\rangle\right)^{\otimes 2}$$

$$= \frac{1}{2\sqrt{2}}(|\uparrow\uparrow\rangle + |\downarrow\downarrow\rangle)\left(|0_{2^m,\Theta}\rangle|0_{2^m,\Theta}\rangle + |1_{2^m,\Theta}\rangle|1_{2^m,\Theta}\rangle\right)$$

$$+ \frac{1}{2\sqrt{2}}(|\uparrow\uparrow\rangle - |\downarrow\downarrow\rangle)\left(|0_{2^m,\Theta}\rangle|0_{2^m,\Theta}\rangle - |1_{2^m,\Theta}\rangle|1_{2^m,\Theta}\rangle\right)$$

$$+ \frac{1}{2\sqrt{2}}(|\uparrow\downarrow\rangle + |\downarrow\uparrow\rangle)\left(|0_{2^m,\Theta}\rangle|1_{2^m,\Theta}\rangle + |1_{2^m,\Theta}\rangle|0_{2^m,\Theta}\rangle\right)$$

$$+ \frac{1}{2\sqrt{2}}(|\uparrow\downarrow\rangle - |\downarrow\uparrow\rangle)\left(|0_{2^m,\Theta}\rangle|1_{2^m,\Theta}\rangle - |1_{2^m,\Theta}\rangle|0_{2^m,\Theta}\rangle\right). \tag{S32}$$

Then after a joint Bell measurement being conducted on the two atoms, the two light modes become entangled. Assuming that the result of the joint Bell measurement is $(|\uparrow\uparrow\rangle + |\downarrow\downarrow\rangle)/\sqrt{2}$, then the corresponding entangled optical state generated is $|0_{2^m,\Theta}\rangle|0_{2^m,\Theta}\rangle + |1_{2^m,\Theta}\rangle|1_{2^m,\Theta}\rangle$ (unnormalized). Then the assumption is made that there is no loss through the fiber channel (the loss will be considered later) and the syndrome measurement is omitted at the moment. According to Section 2.3, after the interaction of the optical state and the suitably prepared atoms in cavities, the atom-light state becomes

$$\frac{1}{2}|\uparrow\uparrow\rangle(|0_{2^m,\Theta}\rangle|0_{2^m,\Theta}\rangle + |1_{2^m,\Theta}\rangle|1_{2^m,\Theta}\rangle)$$

$$+\frac{1}{2}|\uparrow\downarrow\rangle(|0_{2^m,\Theta}\rangle|1_{2^m,\Theta}\rangle + |1_{2^m,\Theta}\rangle|0_{2^m,\Theta}\rangle)$$

$$+\frac{1}{2}|\downarrow\uparrow\rangle(|1_{2^m,\Theta}\rangle|0_{2^m,\Theta}\rangle + |0_{2^m,\Theta}\rangle|1_{2^m,\Theta}\rangle)$$

$$+\frac{1}{2}|\downarrow\downarrow\rangle(|1_{2^m,\Theta}\rangle|1_{2^m,\Theta}\rangle + |0_{2^m,\Theta}\rangle|0_{2^m,\Theta}\rangle)$$

$$=\frac{1}{2}(|\uparrow\uparrow\rangle + |\downarrow\downarrow\rangle)(|0_{2^m,\Theta}\rangle|0_{2^m,\Theta}\rangle + |1_{2^m,\Theta}\rangle|1_{2^m,\Theta}\rangle)$$

$$+\frac{1}{2}(|\uparrow\downarrow\rangle + |\downarrow\uparrow\rangle)(|0_{2^m,\Theta}\rangle|1_{2^m,\Theta}\rangle + |1_{2^m,\Theta}\rangle|0_{2^m,\Theta}\rangle). \quad (S33)$$

Then after discriminating the optical codewords in Alice and Bob's hands, their atomic states finally get entangled. For the case that there is more than one elementary station, a Bell measurement at every $ES_2$ is needed to finally get the states in Alice and Bob's hands entangled, similar to what was mentioned in Section 3.1.

If the photon loss caused by the AD channel is included, the final state of the atoms will remain unchanged whenever we do the Bell measurements on the two atoms in the corresponding elementary station. Thus, the total fidelity is the same as what was derived in the main text.